\documentclass[sigconf]{acmart}
\usepackage{xcolor}
\newcommand{\hlcb}[2]{%
  \begingroup
  \setlength{\fboxsep}{0pt}
  \raisebox{0pt}[0.9ex][0.5ex]{\fcolorbox{#1!80!black}{#1!20}{\strut #2}}%
  \endgroup
}

\usepackage{listings}

\definecolor{narrative}{HTML}{f1bfb3}
\definecolor{investigation}{HTML}{c8d0c8}
\definecolor{reflection}{HTML}{b2cee6}
\definecolor{intergration}{HTML}{dcd6f1}

\AtBeginDocument{%
  }

\acmConference[IUI '26]{ACM Conference on Intelligent User Interfaces}{March 23-26, 2026}{Paphos, Cyprus}

\acmSubmissionID{4247}

\usepackage{multirow}
\usepackage{booktabs}
\usepackage[colorinlistoftodos]{todonotes}

\begin{document}

\begin{teaserfigure}
  \centering
  \includegraphics[width=0.85\textwidth]{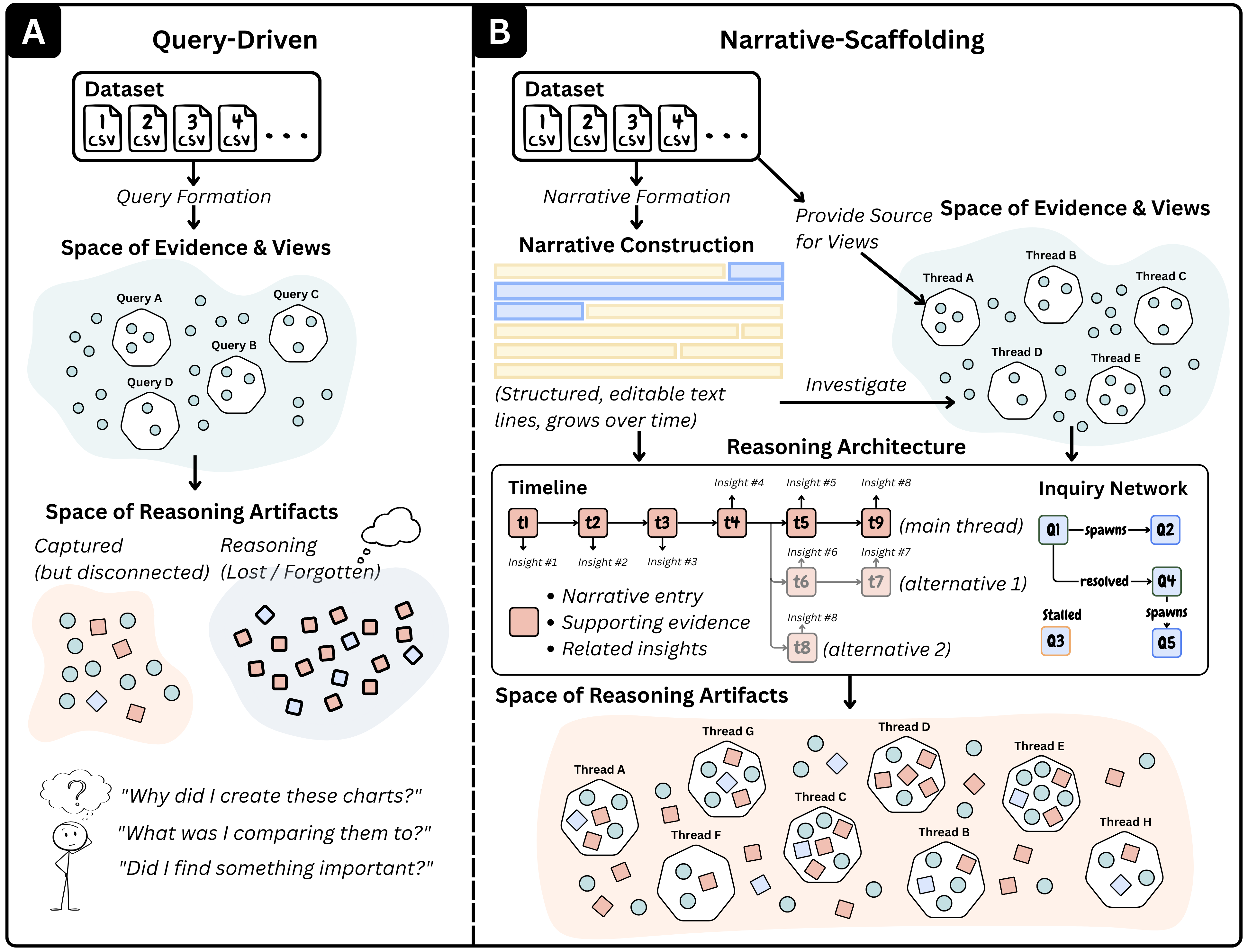}
  \vspace{-0.8em}
  \caption{\textbf{Narrative Scaffolding is a framework for data-driven sensemaking that treats narrative construction as the primary interface for exploration and reasoning.} Unlike traditional query-driven analysis \textbf{(A)}, where users formulate queries to generate disconnected views and reasoning remains implicit, Narrative Scaffolding \textbf{(B)} externalizes evolving reasoning through structured narrative text. As users write, the system generates semantically aligned visualizations, captures insights with provenance, and tracks inquiry evolution. All reasoning artifacts (narrative entries, supporting evidence, and inquiry networks) are organized within a single, connected space, enabling users to revisit, branch, and refine their thinking throughout analysis. }
  \label{fig:teaser}
\end{teaserfigure}


\title{Narrative Scaffolding: A Narrative-First Framework for Data-Driven Sensemaking}

\author{Oliver Huang}
\orcid{0009-0007-1585-1229}
\affiliation{%
  \institution{University of Toronto}
  \country{Toronto, Canada}
}
\email{oliver@cs.toronto.edu}

\author{Muhammad Fatir}
\affiliation{%
  \institution{University of Toronto}
  \country{Toronto, Canada}
}
\email{m.fatir@mail.utoronto.ca}

\author{Steven Luo}
\affiliation{%
  \institution{University of Toronto}
  \country{Toronto, Canada}
}
\email{steventian.luo@mail.utoronto.ca}

\author{Sangho Suh}
\orcid{0000-0003-4617-5116}
\affiliation{%
  \institution{Allen Institute for AI}
  \country{Seattle, USA}
}
\email{sanghos@allenai.org}

\author{Hari Subramonyam}
\affiliation{%
  \institution{Stanford University}
  \country{Stanford, USA}
}
\email{harihars@stanford.edu}

\author{Carolina Nobre}
\orcid{0000-0002-2892-0509}
\affiliation{%
  \institution{University of Toronto}
  \country{Toronto, Canada}
}
\email{cnobre@cs.toronto.edu}

\renewcommand{\shortauthors}{Narrative Scaffolding}

\begin{abstract}
When exploring data, analysts construct narratives about what the data means by asking questions, generating visualizations, reflecting on patterns, and revising their interpretations as new insights emerge. Yet existing analysis tools treat narrative as an afterthought, breaking the link between reasoning, reflection, and the evolving story from exploration. Consequently, analysts lose the ability to see how their reasoning evolves, making it harder to reflect systematically or build coherent explanations. To address this gap, we propose \texttt{Narrative Scaffolding}, a framework for narrative-driven exploration that positions narrative construction as the primary interface for exploration and reasoning. We implemented this framework in a system that externalizes iterative reasoning through narrative-first entry, semantically aligned view generation, and reflection support via insight provenance and inquiry tracking. In a within-subject study ($N=20$), we demonstrated that narrative scaffolding facilitates broader exploration, deeper reflection, and more defensible narratives. An evaluation with visualization literacy experts ($N=6$) confirmed that the system produced outputs aligned with narrative intent and facilitated intentional exploration.


\end{abstract}

\ccsdesc[500]{Human-centered computing~Visual analytics}
\ccsdesc[500]{Human-centered computing~Interactive systems and tools}
\ccsdesc[500]{Human-centered computing~Natural language interfaces}

\keywords{Narrative-Driven Exploration, Narrative Scaffolding, Cognitive Offloading, Visual Analytics, Data Storytelling, AI-Assisted Analytics, Conversational Data Analysis, Insight Provenance, Natural Language Interfaces, Analytical Reasoning}


\maketitle

\section{Introduction}


Exploring data is fundamentally a sensemaking process that involves practicing \textit{interpretive reasoning} throughout: in which analysts continuously form and revise assumptions in response to evidence~\cite{berret2024iceberg, grolemund2014cognitive}. Analysts rarely begin from blank slates~\cite{kandel2012enterprise, wongsuphasawat2019goals}. Instead, they arrive with assumptions about the data (e.g., \textit{"economic growth drives housing demand"}) that shape initial inquiry, such as causal hypotheses (e.g., \textit{"higher income areas will show greater price stability"}), exploratory questions (e.g., \textit{"which boroughs deviate from this pattern?"}), or domain-driven expectations (e.g., \textit{"urban centers tend to recover faster after downturns"})~\cite{rule2018exploration, heuer1999psychology, fu2025dataweaver}. In other words, data exploration entails a cycle of \textit{interpretive reasoning}: analyzing information to extract meaning, questioning or refining these assumptions, and developing evidence-based explanations that guide the course of exploration.


However, this cycle of \textit{interpretive reasoning} often remains implicit~\cite{north2011analytic, dou2009recovering}. Analysts may hold assumptions in memory, pursue lines of inquiry without articulating their rationale, or shift interpretations without documenting the changes that led to them~\cite{lipford2010helping}. When reasoning stays internal, assumptions can harden while still being unverified. Analysts may explore too narrowly, following a single line of inquiry without considering alternatives, or they may selectively seek evidence that confirms initial hunches while overlooking contradictions~\cite{nickerson1998confirmation, wall2017warning}. As exploration deepens, the cognitive burden increases: earlier insights are forgotten, the rationale behind interpretive shifts fades, and promising questions get abandoned without resolution. Without mechanisms to externalize and revisit their evolving thinking, analysts struggle to maintain awareness of how their interpretations have developed, what alternatives they've considered, or which questions remain unresolved.

Constructing narrative offers a structure for making this interpretive reasoning explicit. By structuring questions, interpretations, and evidence into narrative form,  internal thought becomes tangible statements that can be inspected, revised, and linked to supporting evidence~\cite{murray2003narrative, scaife1996external, liu2010mental}. This externalization creates opportunities for reflection, allowing analysts to trace how interpretations evolved, compare competing explanations, and identify gaps in their reasoning~\cite{srinivasan2018augmenting, islam2024datanarrative}. 



However, most analytic tools do not support analysts' narrative as a scaffold for \textit{interpretive reasoning}. They commonly treat narrative as a static output, summarizing finished actions or presenting isolated facts~\cite{keith2022design, wootton2024charting}. Even with recent advances in AI-assisted data analysis that integrate semantics, natural language, and visualization to support more human-centered exploration~\cite{cooperative_setlur, guo2024talk2data, fu2025dataweaver, setlur2025sigmod}, existing systems remain largely reactive: generating visualizations on demand, assembling analysts' narratives retrospectively from logs~\cite{lisnic2025plume, wang2025jupybara, zhao2021chartstory}, and responding to explicit prompts rather than participating in meaning construction as it unfolds~\cite{onTheFly, hindsight, madanagopal2019analytic}. This keeps exploration (data interaction) separate from interpretation (meaning construction)~\cite{tory2021finding, gap_batch}, leaving interpretive reasoning invisible and unsupported~\cite{designProbe, sun2022erato}. While provenance systems document interaction histories, they capture what analysts do but not why. They lack the narrative logic that connects observations into coherent interpretations~\cite{ragan2015characterizing, north2011analytic}. Without structured externalization, analysts often face difficulties engaging in systematic reflection~\cite{dou2009recovering, tory2004human}, which can lead to premature conclusions or leave assumptions unexamined~\cite{nickerson1998confirmation, wall2017warning, confirmation2025}. When reasoning remains implicit, decisions lack the traceability needed for reproducibility and accountability~\cite{gotz2009characterizing, ragan2015characterizing}, which are crucial in fields like journalism~\cite{fu2023more} and intelligence analysis~\cite{saletta2020role}. Ultimately, current approaches limit the ability to share, validate, and build upon insights~\cite{willett2011commentspace, mathisen2019insideinsights}. 

This work addresses these shortcomings by exploring a new interaction paradigm that scaffolds \textit{interpretive reasoning} as it unfolds. By treating narrative construction as the primary interface of the sensemaking process, our approach makes evolving assumptions explicit, surfaces contradictions, and preserves reasoning paths that are otherwise lost. To achieve this, we draw on insights from theories in narrative psychology~\cite{murray2003narrative}, external cognition~\cite{scaife1996external, liu2010mental}, and constructionist learning~\cite{harel1991constructionism, bartram2021untidy}, which collectively emphasize how externalized narratives can serve both as a cognitive medium for reasoning and as an artifact that preserves its evolution.

Grounded in these perspectives, we propose \texttt{Narrative Scaffolding}: a framework that repositions narrative construction as the central interface for data exploration and reasoning. Unlike traditional tools that log low-level interactions or assemble stories retrospectively, \texttt{Narrative Scaffolding} treats interpretation as a dynamic, evolving process that must be externalized and structured to support systematic reasoning. We instantiate this framework in a system that enables analysts to write narrative sentences as first-class analytic actions, with each sentence bidirectionally linked to visual evidence and tracked questions. The system generates semantically-aligned visualizations from natural language, captures insight provenance through a navigable timeline, tracks inquiry evolution through an interactive board, and detects opportunities for branching when reasoning diverges.

In summary, our contributions include:

\begin{itemize}
    \item \textbf{Conceptual framework}: A \texttt{Narrative Scaffolding} framework derived from a formative study that positions narrative structure as the backbone of sensemaking, linking high-level goals, evolving hypotheses, insights, and visual evidence. It emphasizes ongoing reflection, integration of new insights, and preservation of reasoning paths throughout exploration.

    \item \textbf{System instantiation}: An interactive system that operationalizes this framework by making narrative construction the central interface through which analysts pose questions, connect evidence, and develop explanations.
    
    \item \textbf{Empirical evidence}: Results from a two-part evaluation: (1) a within-subject study (N=20) demonstrating that \texttt{Narrative Scaffolding} improves workflow efficiency, supports deeper reflection, encourages broader exploration, and produces more coherent and defensible narratives compared to reactive AI-assisted analysis; and (2) a system-level evaluation (N=6) examining the quality and reliability of AI-generated outputs across diverse analytical scenarios.
\end{itemize}

\section{Related Work}


Understanding how to support interpretive reasoning during data exploration requires examining both how people naturally make sense of information and how existing systems attempt to capture this process. We review prior work through three lenses: the cognitive role of narrative in reasoning, system approaches to capturing analytical provenance and interpretation, and how LLM integration shapes these processes through mixed-initiative interaction.

\subsection{Narrative in Interpretive Reasoning}

Narrative psychology posits that humans are fundamentally wired to make sense of information through constructing coherent narratives to connect events, observations, and meaning~\cite{murray2003narrative, murray2015narrative}. This narrative mode of thought serves not merely as a communication device but as a cognitive tool for organizing experience and building understanding. Sensemaking research supports this view, showing how people iteratively generate and revise interpretations as they build mental models~\cite{pirolli2005sensemaking, klein2007dataframe}, structuring their understanding around storylines that link goals, observations, and conclusions~\cite{dervin1998sense}.

This narrative mode of reasoning manifests across diverse analytical contexts. In journalism, reporters develop story angles that guide their investigation and evidence gathering~\cite{attfield2003information}. In scientific inquiry, researchers frame hypotheses narratively, iterating on explanatory models as new data emerge~\cite{padian2018narrative, dahlstrom2014using}. In data analysis, analysts similarly develop and iterate on storylines for organizing complex information and guiding exploratory decisions~\cite{srinivasan2018augmenting, islam2024datanarrative, fu2023more}. Across these domains, narrative functions as an active reasoning scaffold rather than merely a final output format.

Recent empirical studies validate this narrative-driven reasoning in data exploration. Even without explicit tool support, users spontaneously construct storylines during analysis~\cite{srinivasan2018augmenting}. Observational research shows analysts following interpretive arcs, revisiting assumptions as evidence emerge~\cite{islam2024datanarrative}, and beginning with specific narratives that guide their data exploration~\cite{fu2025dataweaver}. These findings confirm that reasoning unfolds through storyline construction during active exploration, not merely during final communication~\cite{progressiveInsight, zhao2021chartstory}.

Building on these theoretical foundations, our narrative scaffolding framework treats storyline construction as the primary reasoning interface throughout exploration, supporting narrative writing as an active, evolving process of sensemaking.

\subsection{Capturing Analytic Reasoning}

Supporting analytic reasoning requires capturing not just what analysts do, but why they do it and how their understanding evolves. Provenance tracking addresses this by recording the history of how analytical conclusions were reached, capturing the sequence of actions, decisions, and intermediate states that led to the final result~\cite{heer2008graphical}. In nonlinear, iterative exploration, provenance provides a structured record supporting reproducibility, error recovery, collaboration, and accountability~\cite{ragan2015characterizing, north2011analytic, huang2025vistruct, gotz2009characterizing, mathisen2019insideinsights, camisetty2018enhancing}. However, achieving these goals depends critically on what information can actually be captured.

Ragan et al.~\cite{ragan2015characterizing} proposed a foundational framework categorizing provenance into five types: data, visualization, interaction, insight, and rationale. While the first three can be captured through system instrumentation, the latter two involve internal reasoning that is inherently more challenging to observe~\cite{dou2009recovering, north2006toward, willett2011commentspace}. Current systems excel at the observable levels. Workflow-based systems like VisTrails~\cite{callahan2006vistrails} model analysis steps in structured pipelines, though they impose rigid workflows less suited to open-ended exploration. Interactive tracking systems~\cite{heer2008graphical, cutler2020trrack, gratzl2016visual} dynamically log actions and states, while recent approaches like ProvenanceWidgets~\cite{widgets} and ViStruct~\cite{huang2025vistruct} embed tracking directly into interfaces. However, these systems primarily capture observable behaviors while providing limited support for representing underlying reasoning.

To capture insight-level provenance, systems like Calliope~\cite{shi2020calliope} support explicit argument construction by linking claims to data evidence, while CZSaw~\cite{kadivar2009capturing} enables analysts to annotate findings and attach rationale to visual states. However, these approaches require users to explicitly articulate insights at specific moments, creating friction between ongoing exploration~\cite{ragan2015characterizing}. Insights are often documented as static snapshots instead of representing the dynamic nature of evolving thought processes~\cite{north2011analytic}.

This leaves a fundamental gap in capturing interpretive reasoning as it unfolds. Provenance systems excel at documenting what users do but struggle with why they do it. This interpretive evolution (the narrative construction of meaning) is central to how analysts actually reason. When systems cannot capture or support this narrative evolution, reasoning processes remain implicit or fragmented, making it challenging to revisit rationale and trace interpretive evolution~\cite{tory2004human, north2006toward, progressiveInsight}. Our approach addresses this gap by enabling systems to capture insight-level provenance as reasoning naturally unfolds through writing.

\subsection{LLM-Assisted Data-Driven Sensemaking}

Complementary to narrative storyline construction and systematic provenance tracking, mixed-initiative analysis systems address interpretive reasoning through human-AI collaboration~\cite{fu2025dataweaver, lisnic2025plume, latif2021kori, wang2025jupybara}. These systems shift from passive tools to collaborative partners that generate views and respond to natural language queries during exploration.

Current approaches can be categorized into three main interaction paradigms. Conversational interfaces enable dialogue-based exploration where users pose questions and refine queries through natural language~\cite{setlur2016eviza, gao2015datatone, narechania2020nl4dv, obeid2020chart, sultanum2024instruction}. Structured reasoning systems support explicit argument construction through claim-evidence linking and branching logic paths~\cite{shi2020calliope, jiang2023graphologue}, while text-visualization integration approaches facilitate coherence between textual narratives and visual elements~\cite{sultanum2021leveraging, zhu2022crossdata, latif2021kori, pluto}. A key affordance is that text serves as a more expressive interaction medium, enabling iterative communication between users and systems that goes beyond traditional point-and-click interfaces.

A significant focus has emerged on AI-assisted content generation during analysis. These systems leverage LLMs to automatically produce summaries, story fragments, or visual descriptions based on user actions and analytic traces~\cite{fu2025dataweaver, wang2025jupybara, sultanum2023datatales, lisnic2025plume, wang2019datashot}. The effectiveness of these content generation approaches depends on factors such as the sophistication of natural language processing, the quality of AI-generated insights, and the seamless integration of human input with system responses~\cite{li2024we}. Our framework shifts from generating narratives reactively to scaffolding them proactively, treating narrative construction as the primary interface for reasoning throughout exploration.

\section{Formative Studies}
\label{sec:formative}

To understand how interpretive reasoning unfolds during data exploration and 
where current tools create friction, we conducted a formative study that organized these issues into four interpretive challenges. We surveyed \textit{48} analysts to validate the prevalence of challenges and conducted follow-up interviews with \textit{5} participants to examine how reasoning strategies evolve. A preliminary version of these findings was reported 
in~\cite{huang2025toward}. Below, we detail the procedure for survey and interview studies and describe the identified challenges that informed our design goals.

\subsection{Procedure}
\subsubsection{Formative Survey ($N=48$)}
We surveyed 48 data practitioners ($F_1$--$F_{48}$) who frequently engage in data exploration. Participants were recruited through Prolific~\cite{prolific2024} and compensated at the rate of \$16 USD/hr. The sample (26 male, 22 female, ages 21--55) included data analysts ($n=16$), managers ($n=7$), healthcare professionals ($n=6$), and sales consultants ($n=3$). This ensured diversity across professional domains, capturing a broad range of reasoning contexts. The survey combined Likert items with semi-structured open-ended prompts, probing how participants initiated exploration, adapted their thinking when encountering unexpected findings, managed analytical complexity, and revisited prior insights. (see supplementary materials for the full protocol)

Responses revealed common behavioral patterns: managing multiple exploration paths through workarounds, externalizing reasoning outside the analytic environment, and restarting from scratch when interpretations diverged. To further investigate how interpretive reasoning unfolds over time and what triggers specific analytical decisions, we conducted follow-up interviews with five participants.

\subsubsection{Follow-up Interview ($N=5$) and Challenge Formation}

We recruited five participants from the survey ($F'_1$--$F'_5$) who have experience in narrative-driven analytical workflows and typically using mixed-initiative analytic tools. Each 30--45 minute interview (\textit{M=39.2}) was compensated at a rate of \$20 USD/hr and followed a semi-structured protocol where participants walked through recent analysis sessions, describing how they initiated exploration, articulated evolving goals, and managed reasoning shifts in response to new evidence. (See supplementary materials for the full protocol.)

The interviews revealed rich insights which directly informed our design goals. Open-ended survey responses and interview transcripts were analyzed using inductive thematic analysis~\cite{nowell2017thematic}. Table~\ref{tab:challenge_strategy_goal} summarizes the four identified challenges (\textbf{C1--C4}), observed behaviors (\textbf{B1--B9}), and corresponding design goals (\textbf{DG1--DG4}).
 
\begin{table*}[]
\caption{Summary of user challenges, observed barriers, design goals, and system features addressing interpretive reasoning in data exploration.}
\label{tab:challenge_strategy_goal}

\begin{tabular}{p{0.22\textwidth}|p{0.25\textwidth}|p{0.22\textwidth}|p{0.22\textwidth}}
\toprule
\textbf{C}hallenges & Observed \textbf{B}arriers & \textbf{D}esign \textbf{G}oals & System Features \\
\toprule

\multirow{3}{0.22\textwidth}{\textbf{C1}: Unsupported transition from exploration intent to evidence (Sec~\ref{sec:challenge-1})}
& \textbf{B1}: Exploration intentions lack detail
& \multirow{3}{0.22\textwidth}{\textbf{DG1}: Support fluid transitions from vague inquiries to evidence-grounded insights}
& Semantic-aligned views \\ \cline{2-2}\cline{4-4}
& \textbf{B2}: Generated views become disconnected
&
& Bidirectional narrative construction \\ \cline{2-2}
& \textbf{B3}: Reliance on external note tools
&
& \\ \midrule

\multirow{3}{0.22\textwidth}{\textbf{C2}: Unmanaged inquiry and reflective continuity (Sec~\ref{sec:challenge-2})}
& \textbf{B4}: Interpretations not explicitly captured
& \multirow{3}{0.22\textwidth}{\textbf{DG2}: Externalize inquiry intent and maintain continuity of reflection}
& Insight timeline \\ \cline{2-2}
& \textbf{B5}: Limited reflective review
&
& \\ \cline{2-2}\cline{4-4}
& \textbf{B6}: Reasoning paths left unresolved
&
& Inquiry board \\ \midrule

\multirow{2}{0.22\textwidth}{\textbf{C3}: Disrupted divergence in exploratory reasoning (Sec~\ref{sec:challenge-3})}
& \textbf{B7}: Restarts instead of branching
& \multirow{2}{0.22\textwidth}{\textbf{DG3}: Support nonlinear exploration and alternative reasoning}
& Narrative branching \\ \cline{2-2}
& \textbf{B8}: Hard to compare alternative paths
& \\ \midrule

\multirow{1}{0.22\textwidth}{\textbf{C4}: Unstructured notes lacking formalization (Sec~\ref{sec:challenge-4})}
& \textbf{B9}: Difficult synthesis into analysis
& \textbf{DG4}: Support structured and defensible narratives
& Story Integration \\
\\
\bottomrule
\end{tabular}
\end{table*}

\subsection{Challenges}

\subsubsection{\textbf{Challenge 1: Unsupported Transition from Intent to Evidence}}
\label{sec:challenge-1}

Prior work shows that analysts often begin exploration with tentative or underspecified intentions, yet most analytic systems demand premature precision or rigid query formulation~\cite{pirolli2005sensemaking, heer2008graphical, ragan2015characterizing}. In our survey, \textit{39} of \textit{48} participants reported that their early exploration goals lacked detail (\textbf{B1}). $F_{12}$ reflected that \textit{"It gives me something too specific when I'm still just exploring"}. This mismatch limits how users pursue their curiosity or flexibly define intentions.

As exploration progressed, \textit{41} of \textit{48} reported that generated views became disconnected from their evolving questions (\textbf{B2}). $F_{21}$ explained, \textit{"I take screenshots when something surprises me, but later I forget why it mattered."} Without mechanisms to anchor interpretive context to evidence, visualizations quickly lose their meaning, making them less valuable to revisit.

To manage the situation, participants (\textit{36} of \textit{48}) offloaded their reasoning into external note-taking tools, such as slides or documents (\textbf{B3}). $F_{6}$ noted, \textit{"I'll write down an idea like 'maybe this drop is due to [factor],' but that note just lives separately from the data."} These workarounds highlight how reasoning fragments drift apart from evidence, creating temporal degradation. As $F'_{2}$ reflected, \textit{"I usually remember my conclusions, but not the reasoning that led me there."} This illustrates how outcomes are retained while the reasoning behind them fades, making it harder to revisit or refine past analyses.

\subsubsection{\textbf{Challenge 2: Unmanaged Inquiry and Reflective Continuity}}
\label{sec:challenge-2}

As exploration unfolded, many interpretations that participants formed were never made explicit within their analytic environment. From the survey, \textit{44} of \textit{48} participants acknowledged that their reasoning often remained undocumented during the process (\textbf{B4}). $F_{14}$ explained, \textit{"I usually have an idea in my head, if I don't write it down, it just disappears as I move forward."}  When interpretations are not explicitly captured, they often fade from memory as analysis progresses, which limits the continuity of inquiry and weakens opportunities for reflection. As $F'_{3}$ explained, \textit{"I sometimes can't get back to what I thought earlier, so I just move forward instead."}

Even when participants attempted to revisit their earlier reasoning, the process was limited and inconsistent. \textit{32} of \textit{48} participants reported rarely reviewing past conclusions (\textbf{B5}). $F'_{5}$ described, \textit{"I sometimes go back to see if my conclusion still makes sense, but it's hard to remember the context of why I thought it."} Without preserved context, retrospective reflection becomes fragmented and difficult to sustain~\cite{hullman2022judgment, kale2022discern}.

Finally, reasoning paths often ended unexpectedly as participants shifted focus. \textit{25} of \textit{48} participants acknowledged abandoning questions without resolution (\textbf{B6}). As $F'_{2}$ noted, \textit{"I left one question midway, but I never went back to see if it was still valid."} This lack of continuity results in abandoned questions and forgotten assumptions, making it difficult to trace how reasoning evolved.

\subsubsection{\textbf{Challenge 3: Disrupted Divergence in Exploratory Reasoning}}
\label{sec:challenge-3}

Most analytic tools record actions in a single linear sequence~\cite{gotz2009characterizing, gratzl2016visual}, but reasoning often branches as analysts test alternatives~\cite{pirolli2005sensemaking, north2011analytic}. Participants (\textit{29} of \textit{48}) reported restarting or duplicating their workspace when their thinking shifted (\textbf{B7}). $F_{21}$ explained, \textit{"If I want to try a new idea, I just duplicate the current exploration path and restart another, changing the original feels risky."} Restarting in this way creates a break in continuity: earlier interpretations are severed from the new line of thought, forcing participants to "\textit{manage multiple versions manually}" ($F_{43}$). 

Prior work has shown that a narrative-driven approach can inadvertently reinforce existing assumptions when alternative explanations are not explored~\cite{nickerson1998confirmation, confirmation2025, wall2017warning}. Participants (\textit{33} of \textit{48}) emphasized the importance of considering alternative explanations to maintain a balanced approach to their reasoning. However, a further \textit{23} of \textit{48} described difficulty keeping track of multiple possible directions when their reasoning diverged (\textbf{B8}). $F'_{4}$ reflected, \textit{"When I try two different parallel explanations, I usually lose track of one of them."} Without support to manage alternatives side by side, users struggled to weigh competing possibilities, which could limit consideration of disconfirming evidence.

\subsubsection{\textbf{Challenge 4: Unstructured Notes Lacking Analytical Formalization}}
\label{sec:challenge-4}

While participants extensively captured interpretive thoughts during the exploration, these insights remained fragmented and unstructured, making them difficult to reuse or communicate effectively. Most participants (46 of 48) reported difficulty reconstructing their analytical logic from informal notes and scattered observations (\textbf{B9}). $F_{10}$ noted, \textit{"I sometimes conclude quickly, and later I can't remember why that was in the first place."} Others described how their reasoning fragments lacked structure for meaningful reuse. $F'_{1}$ explained, \textit{"I leave comments in documents, but they never translate back into the final analysis itself."} 

This fragmentation creates a temporal sensemaking problem where insights lose their analytical context over time. $F'_{5}$ remarked, \textit{"I jot things down as I go, but those notes don't connect to show how my thinking evolved."} The challenge lies not merely in communication to stakeholders, but in preserving the analytical coherence needed for iterative reasoning and insight building.

\subsection{Informed Design Goals}\label{sec:goals}
Based on the outlined challenges (\textbf{C1--C4}) from our formative study, we developed four design goals (\textbf{DG1-DG4}) to address the observed barriers (\textbf{B1--B9}). These goals informed our system architecture and aim to scaffold \textit{interpretive reasoning} by externalizing evolving insights, supporting reflection, and enabling flexible narrative exploration.\\

\noindent\textbf{\textit{DG1: Support fluid transitions from vague inquiries to evidence-grounded insights.}}  
To address the difficulty of moving from tentative goals into evidence-grounded analysis (\textbf{C1}), the system should provide semantic-aligned view generation that transforms underspecified starting points into meaningful visualizations (\textbf{B1}). It should also support linked authoring of narrative evidence to keep interpretive insights embedded in visual context rather than separated into external notes (\textbf{B2, B3}), enabling analysts to revisit earlier thoughts with continuity.

\noindent\textbf{\textit{DG2: Externalize inquiry intent and maintain continuity of reflection.}}
To address undocumented interpretations and abandoned questions (\textbf{C2}), the system should provide insight-level provenance that records reasoning alongside actions, allowing users to retrace prior thoughts (\textbf{B4, B5}). An inquiry board should maintain continuity by keeping unresolved questions visible and available for later return (\textbf{B6}), reducing reliance on memory alone.

\noindent\textbf{\textit{DG3: Support nonlinear exploration and alternative reasoning.}} 
To enable divergent reasoning without breaking continuity (\textbf{C3}), the system should support narrative branching and direct comparison of alternatives (\textbf{B7, B8}). By making multiple reasoning paths visible and manageable, the system helps analysts systematically consider competing explanations and broaden their exploration scope.

\noindent\textbf{\textit{DG4: Support structured and defensible narratives.}}
To bridge scattered notes and informal observations into coherent outputs (\textbf{C4}), the system should scaffold narrative construction by maintaining evidence linkages throughout exploration (\textbf{B9}). This ensures final outputs remain defensible and reduces the effort required to reorganize materials for communication.

\section{\texttt{Narrative Scaffolding} Framework}
\label{sec:framework}
\begin{figure*}[ht]
    \centering
    \includegraphics[width=0.95\linewidth]{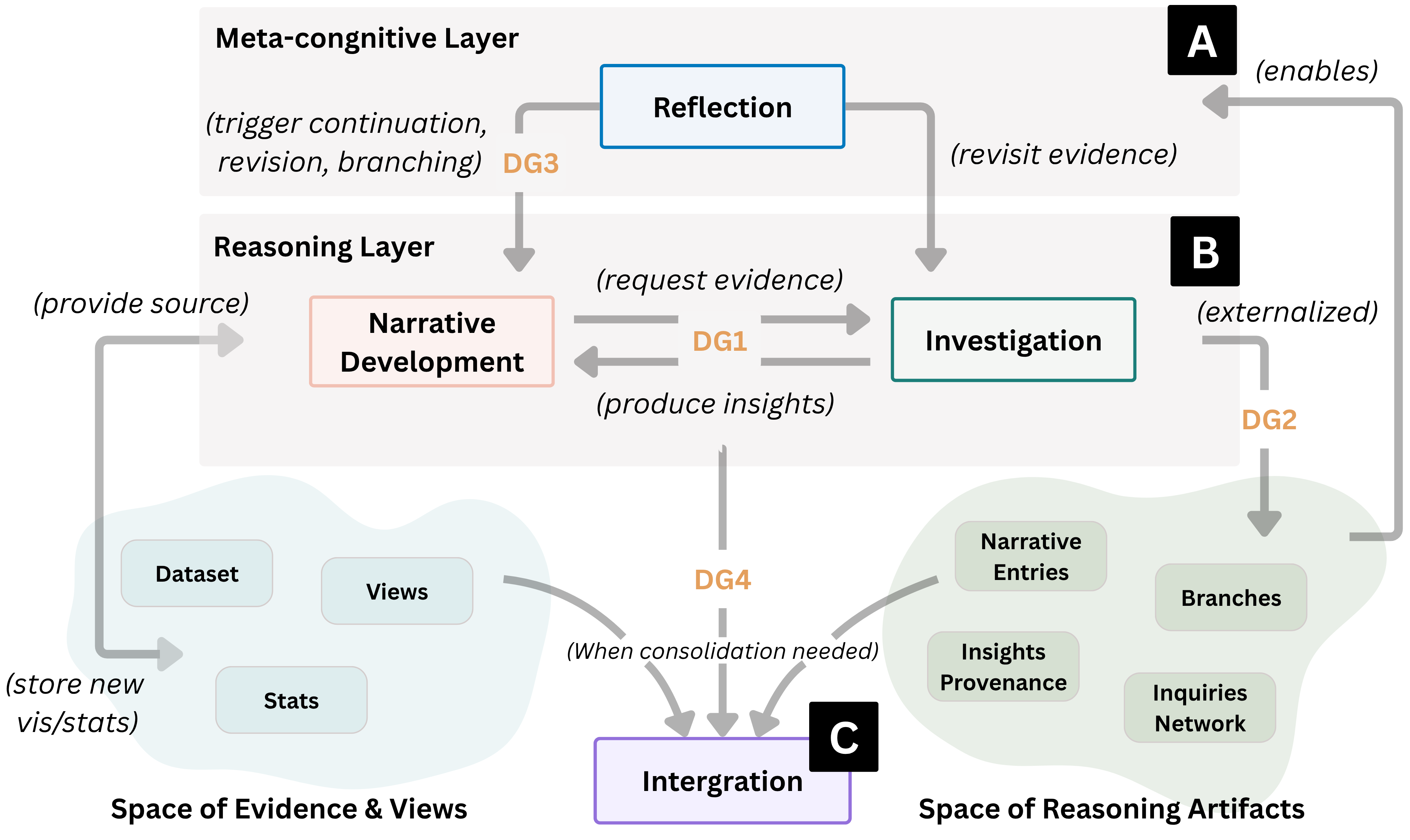}
    \caption{\textbf{The Narrative Scaffolding framework.} Analysts begin with narrative development, where assumptions and questions are externalized in response to the dataset at hand. These narratives guide investigation, producing visualizations and insights grounded in the data. Insights feed back into narrative development as new or revised hypotheses. Reflection occurs throughout, as analysts revisit their externalized reasoning and evidence. Integration consolidates claims, branches, and supporting data into coherent accounts.}
    \label{fig:conceptual-model}
\end{figure*}

\subsection{Stages and Theoretical Grounding}

To connect our established design goals (\textbf{DG1-DG4}) to actual analytic practice --- and use it to ground the design of our system (presented in Sec~\ref{sec:system}), we propose a conceptual model that captures how narrative construction unfolds through distinct stages of \textit{interpretive reasoning}. This model provides a structured view of the cognitive processes our system is designed to support. The model was adapted from two established perspectives on analytic reasoning: the Sensemaking Loop~\cite{pirolli2005sensemaking}, which describes an iterative cycle between searching for information and organizing it into mental structures, and the Data--Frame model~\cite{klein2007data}, which emphasizes how analysts construct, test, and revise interpretive frames in response to new evidence. We also draw on external cognition theory~\cite{scaife1996external} and narrative psychology~\cite{murray2003narrative} to ground how each stage supports reasoning. Building on these foundations, as shown in  Fig.~\ref{fig:conceptual-model}, our adaptation repositions narrative construction as the primary reasoning interface, and maps four stages to our design goals. 

\begin{itemize}  
    \item \textbf{\hlcb{narrative}{Narrative Development}}: Externalizing assumptions as narrative entries that can be revised or branched.

    \item \textbf{\hlcb{investigation}{Investigation}}: Using narrative claims to generate evidence, and using evidence to reshape claims.  

    \item \textbf{\hlcb{reflection}{Reflection}}: Revisiting explicit records of reasoning to check coherence, contradictions, and alternatives.

    \item \textbf{\hlcb{intergration}{Integration}}: Consolidating dispersed reasoning into provisional but defensible accounts.
\end{itemize}  

Each stage operationalizes specific theoretical constructs. \linebreak{}\textbf{\hlcb{narrative}{Narrative Development}} draws on external cognition theory: by externalizing assumptions as written text, analysts transform fleeting thoughts into manipulable artifacts that can be inspected and revised~\cite{scaife1996external, liu2010mental}. \textbf{\hlcb{investigation}{Investigation}} operationalizes the Data-Frame model's emphasis on testing interpretive frames against evidence, where claims and data mutually reshape one another~\cite{klein2007data}. \linebreak{}\textbf{\hlcb{reflection}{Reflection}} embodies the Sensemaking Loop's iterative cycling between bottom-up evidence gathering and top-down schema revision~\cite{pirolli2005sensemaking}. \textbf{\hlcb{intergration}{Integration}} draws on narrative psychology's insight that constructing coherent narratives is itself a mode of reasoning, not merely a post-hoc communication step~\cite{murray2003narrative, bruner1991narrative}.

\subsection{Framework Operation}

\texttt{Narrative Scaffolding} operates as a cyclical reasoning loop structured across two layers. At the \textbf{reasoning layer} (Fig.~\ref{fig:conceptual-model}, B), analysts move between \textbf{narrative development} and \textbf{investigation}. Narrative development externalizes assumptions and emerging claims as narrative entries, turning them into manipulable artifacts rather than fleeting thoughts (\textbf{DG1}). These entries can be revised or branched into alternatives, preserving multiple lines of reasoning rather than collapsing them into a single path (\textbf{DG3}). Narrative entries also drive investigation, generating semantically aligned visualizations that test or expand the claims. Investigation produces insights, which are themselves externalized back into the narrative, ensuring evidence and reasoning remain tightly coupled (\textbf{DG1}).  

At the \textbf{meta-cognitive layer} (Fig.~\ref{fig:conceptual-model}, A), \textbf{reflection} regulates this cycle. Because reasoning has been externalized through narrative entries, insight timelines, and inquiry boards, reflection is no longer reliant on memory. Instead, analysts can revisit earlier claims, inspect supporting evidence, and recognize contradictions or gaps (\textbf{DG2}). In doing so, reflection often sparks additional narrative development, introducing new hypotheses or branches that expand the inquiry (\textbf{DG3}).  

\textbf{Integration} (Fig.~\ref{fig:conceptual-model}, C) occurs when analysts choose to consolidate what has been externalized so far. It utilizes a comprehensive set of externalized reasoning, including active claims, alternative branches, and supporting evidence, to create coherent and defensible accounts (\textbf{DG4}).

Figure~\ref{fig:conceptual-model} highlights this dependency: narrative development externalizes thoughts; reflection acts on those externalizations; investigation attaches evidence to them; and integration consolidates the entire set into coherent narratives. Narrative construction thus enables reasoning to remain visible, traceable, and revisable at every stage. The following section describes how these framework stages are instantiated as concrete system features and interactions in the Narrative Scaffolding system.

\begin{figure*}[ht]
    \centering
    \includegraphics[width=0.95\linewidth]{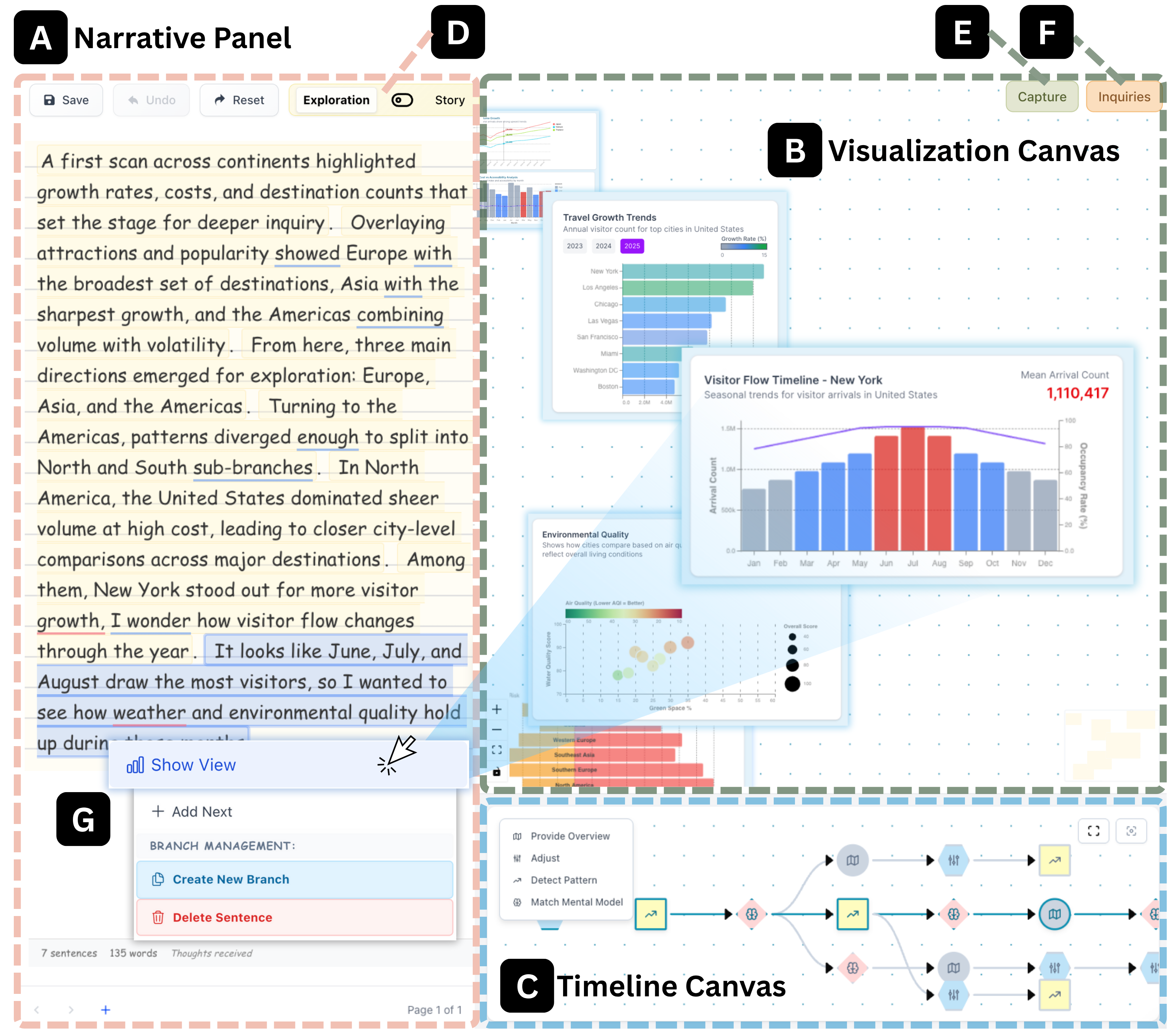}
    \caption{\textbf{The Narrative Scaffolding interface integrates three canvases}. (A, left) The Narrative Panel is the primary authoring space where analysts articulate, revise, and branch reasoning. (B, top right) The Visualization Canvas displays semantically aligned views that respond to narrative claims and capture user interactions as evidence. (C, bottom right) The Timeline Canvas records how insights evolve over time, preserving provenance as a navigable sequence of reasoning steps. Together, these canvases provide a shared workspace where narrative and evidence remain continuously linked.}
    \label{fig:interface}
\end{figure*}

\section{\texttt{Narrative Scaffolding} System}
\label{sec:system}

This section details how our system instantiates the conceptual model (Fig.~\ref{fig:conceptual-model}), which has four core stages: 
\hlcb{narrative}{narrative development}, \hlcb{investigation}{investigation}, \hlcb{reflection}{reflection}, and \hlcb{intergration}{integration}.
Specifically, we describe the mechanisms driving each stage: how the system supports narrative construction and branching~(\ref{sys/construction}), enables investigation through narrative--evidence coupling~(\ref{sys/coupling}), scaffolds multi-level reflection~(\ref{sys/reflection}), and integrates insights into coherent, data-driven stories~(\ref{sys/integration}).


\subsection{\hlcb{narrative}{Narrative Development}: Reasoning Through Construction \& Branching}\label{sys/construction}

\begin{figure*}[ht]
    \centering
    \includegraphics[width=1\linewidth]{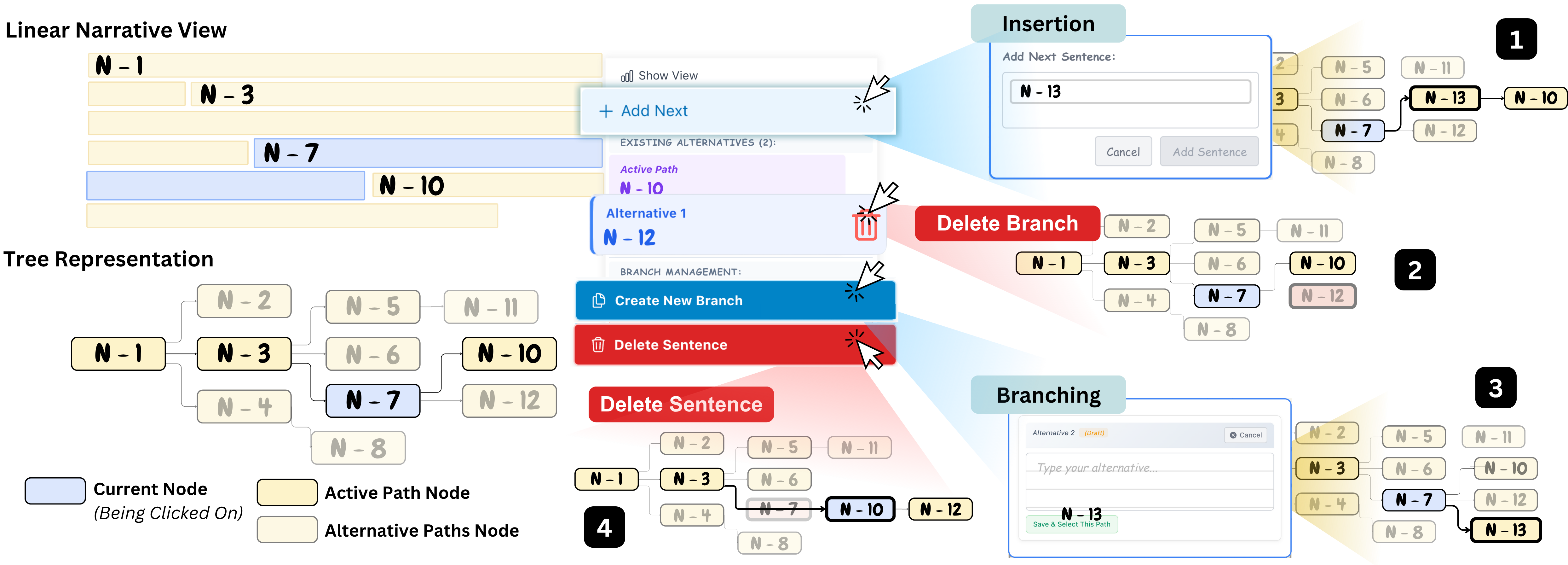}
    \caption{\textbf{Core narrative editing operations in Narrative Scaffolding.} Analysts construct reasoning as editable narrative trees, using the linear narrative view and the branching tree representation. The system supports lightweight operations for (1) inserting new sentences, (3) creating branches for alternative reasoning paths, and (2) deleting sentences or (4) branches while maintaining coherent structure.}
    \label{fig:construction}
\end{figure*}

Narrative construction is the core interface of our system, where reasoning is made explicit and evolves throughout exploration. Narrative text is maintained as a tree-structured sequence of sentences, rather than a fixed linear document, allowing analysts to expand and adapt their stories as new insights emerge. Users shape this structure through four lightweight operations: appending, inserting, updating, and deleting sentences (Figure~\ref{fig:construction}). These operations treat each narrative entry as a manipulable object, enabling claims, questions, and interpretations to be progressively articulated and revised.
When alternative explanations arise, users can initiate \textbf{narrative branching} from any sentence (Figure~\ref{fig:interface}, G), generating a parallel path that inherits all prior context. This design preserves multiple interpretations without collapsing exploration into a single account, supporting both divergence and continuity in reasoning.



\subsection{\hlcb{investigation}{Investigation}: Narrative--Evidence Coupling}\label{sys/coupling}

Investigation is the central stage of narrative scaffolding, where the narrative's intent is transformed into evidence, and that evidence subsequently generates new narrative directions. At this stage, user-authored claims generate semantically aligned visualizations and statistical summaries, while interactions with these views can be captured and integrated back into the narrative as insights. These artifacts keep reasoning linked to their supporting data, and at the same time, provide material for analysts to explore alternatives and formulate new intent.

\subsubsection{\textbf{Semantic-Aligned View Generation}}~\label{sys:pipeline}

A core function of investigation is transforming underspecified narrative intent into meaningful visualizations. To address this, we developed a semantic alignment 
pipeline that matches narrative statements to relevant visualizations through a joint embedding space of narrative propositions and chart specifications (detailed in Appendix Figure~\ref{fig:pipeline}).

The system preprocesses dataset schemas to generate narrative proposition templates covering common analytical patterns (ranking, temporal change, outlier detection) and pairs them with Vega visualization specifications. Both are normalized into semantic tags and embedded alongside pre-generated text-to-SQL queries (see Appendix~\ref{pipeline} for implementation details).

When users write a narrative and select \textit{\hlcb{narrative}{Show View}} 
(Figure~\ref{fig:interface}, G), the system matches their text against 
this joint space, retrieves relevant propositions, and determines whether 
a single chart or dashboard best addresses the claim. For instance, "Asia 
is cheaper but more crowded" triggers a dashboard with cost and crowding 
metrics, while "Porto stands out for affordability" generates a focused 
comparison chart. The system executes corresponding SQL queries and renders 
interactive visualizations, enabling meaningful view generation from vague 
inputs while maintaining semantic consistency.

\subsubsection{\textbf{Interaction-to-Narrative Capture}}

Exploration does not only flow from narrative to evidence, interactions with evidence can also be reformulated back into narrative form. As users interact with charts through actions such as clicks, hovers, or filters, the system records these operations as part of the investigative context. The five most recent interactions, together with the currently active narrative claims, are passed to the LLM to generate candidate narrative statements. When users click \textit{\hlcb{investigation}{Capture}} (Figure~\ref{fig:interface}, E) in the visualization canvas, these candidate insights summarize what the analyst is probing in the data and are presented for capture into the narrative pane. Once accepted, they are appended as narrative sentences and simultaneously logged in the \textit{\hlcb{reflection}{Insight Timeline}}, preserving both the text and its supporting evidence as artifacts.

\subsection{\hlcb{reflection}{Reflection}: Tracing Reasoning Evolution}\label{sys/reflection}
Reflection provides moments to revisit, evaluate, and reframe emerging narratives. In the system, this process is structured across three layers: expression, interpretation, and reasoning. These layers ensure that users can re-engage with their analysis, whether by recalling a single claim, examining a sequence of insights, or assessing the overall trajectory of inquiry.


\subsubsection{\textbf{Sentence-to-View Linking}}

\begin{figure}[tb]
    \centering
    \includegraphics[width=\linewidth]{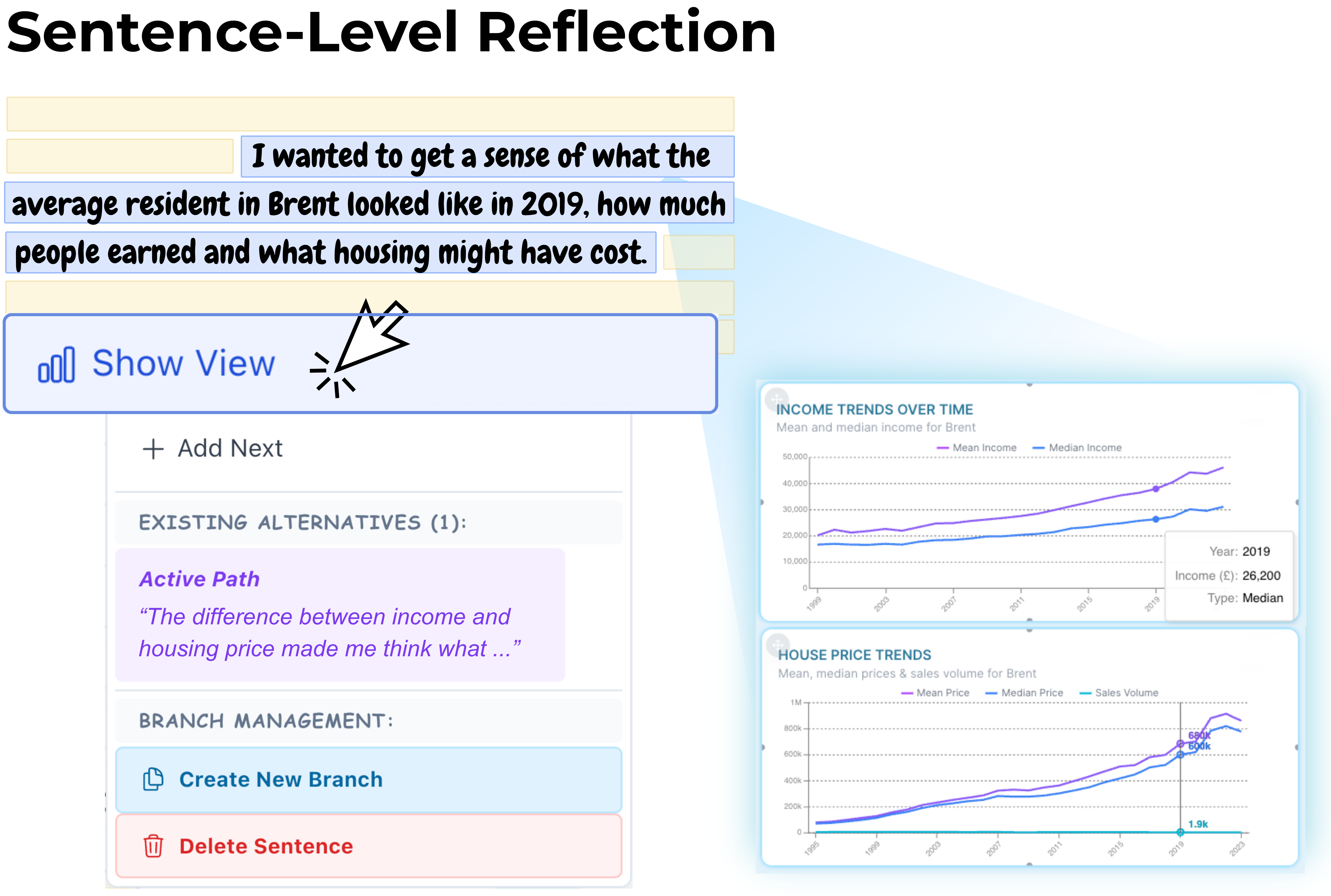}
    \caption{Sentence-level reflection allows analysts to revisit specific narrative statements in context, linking individual sentences to visual evidence for localized reasoning.}
    \label{fig:reflection}
\end{figure}

The narrative panel supports direct recall between narrative and evidence (Figure~\ref{fig:reflection}). Every visualization generated during the investigation is preserved on the canvas and linked to the sentence that introduced it. This linkage is bidirectional: users can recall a visualization from its associated sentence, or select a visualization to highlight the claim it supports. At this level, reflection occurs on a local, sentence-by-sentence basis, providing in-the-moment awareness (i.e., \textit{"Am I making a claim? Have I justified it?"}). This bidirectional recall ensures that evidence and expression remain tightly connected throughout the analysis.

\subsubsection{\textbf{Insight Provenance Timeline}}

\begin{figure*}[!tbp]
  \centering
  \begin{minipage}[b]{0.49\textwidth}
    \includegraphics[width=\textwidth]{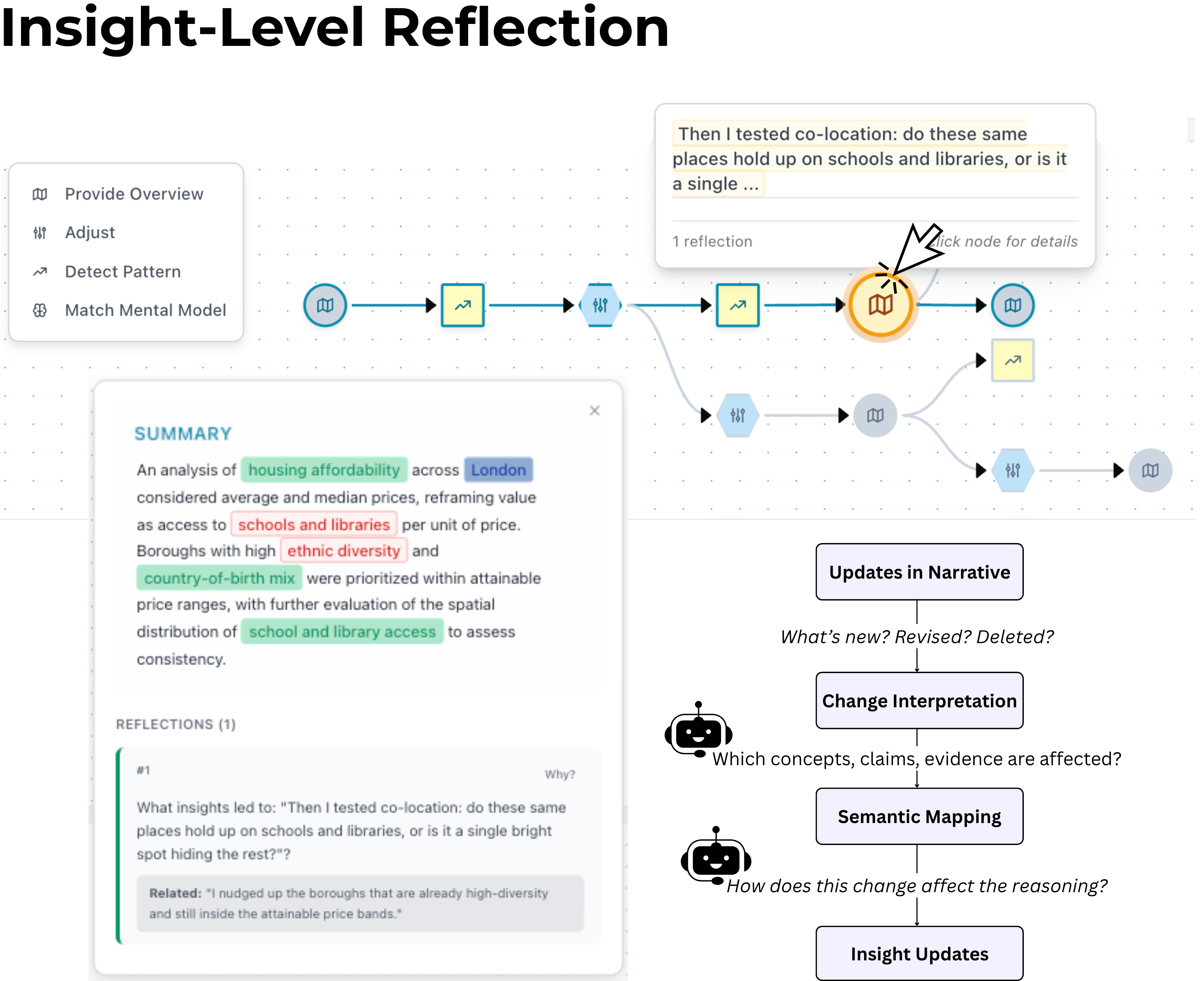}
        \caption{Insight-level reflection captures how edits in the narrative propagate to interpretive changes in the reasoning structure.}
    \label{fig:insight}
  \end{minipage}
  \hfill
  \begin{minipage}[b]{0.49\textwidth}
    \includegraphics[width=\textwidth]{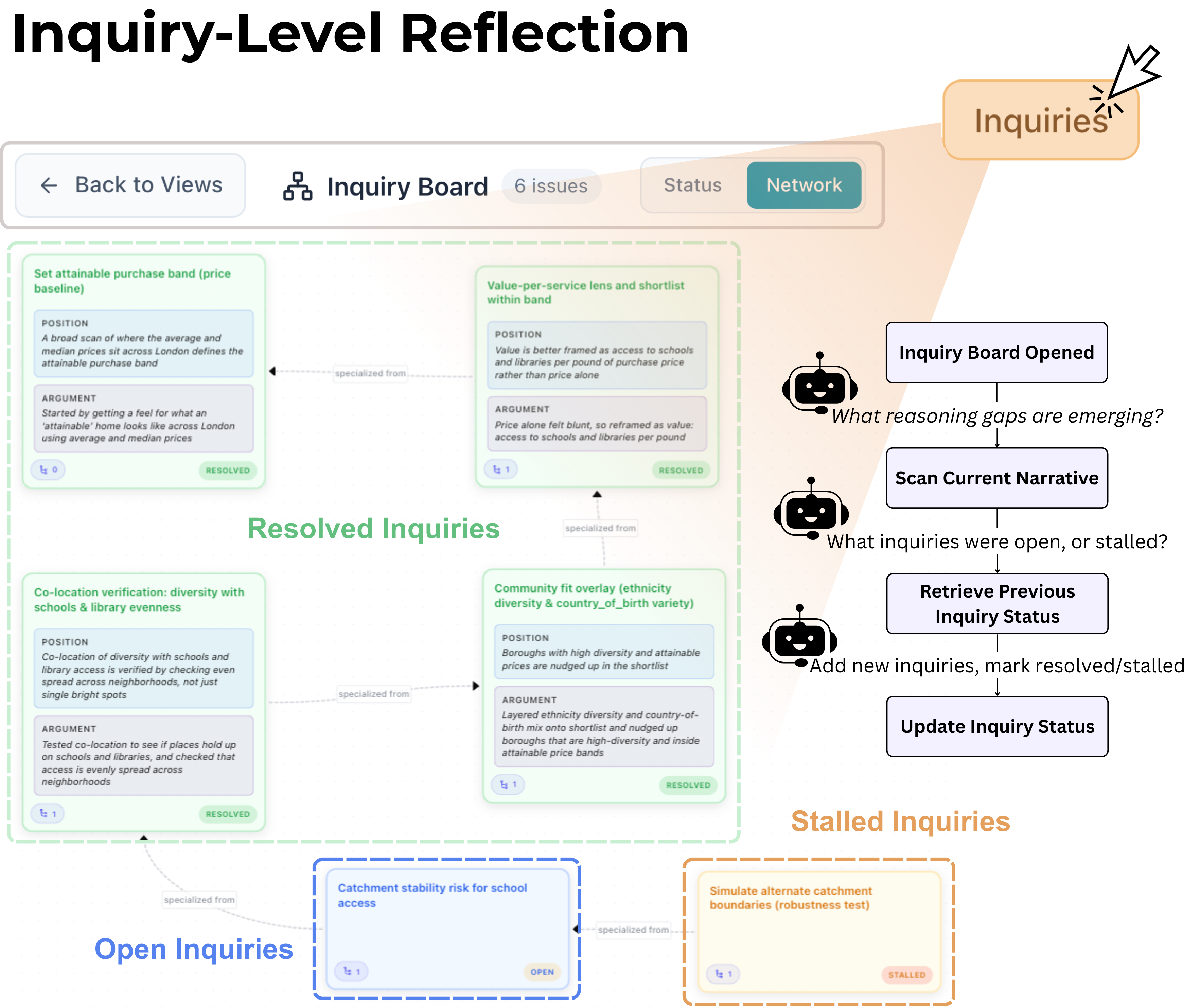}
    \caption{Inquiry-level reflection synthesizes across insights, reviewing and linking how questions evolve and progress over time.}
    \label{fig:inqury}
  \end{minipage}
\end{figure*}

The insight timeline captures how insights emerge and evolve as users write, recording each update as a node (Figure~\ref{fig:insight}). We adapted the Data--Frame taxonomy~\cite{klein2007data, yi_insights_2008} to classify insights into four relationship types describing how claims shift the trajectory of reasoning: \textbf{provide overview}, \textbf{adjust}, \textbf{detect pattern}, and \textbf{match mental model}. The timeline is organized as a tree structure, preserving both the current path and alternative branches.
At this level, reflection focuses on how a user's thinking shifts over time, helping users answer questions such as \textit{"How did I arrive at this claim?"} or \textit{"Has my hypothesis drifted?"}. By visualizing the trajectory of reasoning as a tree, this design provides a navigable record that allows users to revisit and recover the precise state of thought at any point in their exploration.



\subsubsection{\textbf{Inquiry Board}}

The inquiry board captures how inquiries unfold over time (Figure~\ref{fig:inqury}). Inquiries are derived directly from the narrative: when users write hypotheses or intentions to explore, these statements are formalized as issues, while claims provide potential answers. Because exploration is ongoing, many inquiries may remain unanswered or give rise to new ones as paths branch. The board categorizes all current inquiries into three statuses: \textbf{open}, \textbf{resolved}, or \textbf{stalled}, and arranges them in an IBIS graph~\cite{conklin1988gibis, kunz1970issues} to show how issues relate to one another. At this level, reflection shifts from local claims to the broader organization of reasoning. The inquiry board enables users to revisit unresolved issues, re-engage with abandoned threads, and assess whether current claims have sufficient support, preventing forgotten questions from fading silently out of the reasoning process.\\

\noindent{}Together, these layers scaffold narratives through reflection at multiple levels: from local articulation of claims, to tracing how insights evolve, to organizing inquiries into a higher-level reasoning structure.

\subsection{\hlcb{intergration}{Integration}: Synthesis as Data-Driven Story}\label{sys/integration}

\textbf{Integration} reorganizing nonlinear exploration into a single, communicable narrative grounded in the user's reasoning process. 
Triggered by switching to the \textit{\hlcb{intergration}{Story Mode}} (Figure~\ref{fig:interface}, D), the system compiles all narrative branches and identifies key conclusions, decision points, and supported insights across the entire insight timeline. The LLM is prompted with the entire narrative graph structure, including associated inquiry paths and metadata that link each claim to evidence and its corresponding inquiry context. Only insights that are statistically grounded are included to ensure coherence and relevance. The system generates a clean, linear narrative where each paragraph contains synthesized insights linked back to their supporting visualizations.


\subsection{Usage Scenario and System Walk-through}

Imagine Amy, a data journalist, receives an assignment to analyze several travel datasets for spring destination recommendations. 

\textbf{\hlcb{narrative}{Narrative Development} \& \hlcb{investigation}{Investigation}}. She begins with intention and narrative development, drafting an opening claim in the narrative pane about balancing affordability, environmental quality, and crowding. This externalized sentence guides her exploration, and when she clicks \textit{\hlcb{narrative}{Show View}}, the system generates aligned dashboards: cost distributions by region, sustainability rankings, and popularity metrics, that ground her claim in evidence.

As Amy examines the visualizations, she poses a new question about European sustainability premiums compared to Asia, and the system responds with targeted comparisons. Clicking \hlcb{investigation}{\textit{Capture}} saves her observation (\textit{"Asia is 21.5\% cheaper, but 62.8\% more crowded"}) into the narrative, while also attaching the supporting views. The \textit{\hlcb{reflection}{Insight Timeline}} records a new node, ensuring both her claim and the underlying data remain visible in the evolving reasoning process. Her assumptions and observations are captured by the system as externalized artifacts that remain persistently linked to the visual evidence that generated them.

\textbf{\hlcb{reflection}{Reflection}}. Drilling down from regional to city-level patterns, Amy captures insight \textit{"Porto stands out for budget appeal among mid-size cities"}, saved to the narrative with linked cost distributions and a timeline node. As Amy continues building her narrative around multiple destinations, she realizes she can't recall whether Porto's low cost reflected seasonal pricing or year-round affordability. To verify this detail, she needs to revisit earlier reasoning. She reopens her Porto claim with \textit{\hlcb{investigation}{Show View}}, inspects the attached visuals, and notices a weakness in environmental quality that contradicts her initial assumption. She edits the sentence to qualify the recommendation (\textit{"Porto offers budget appeal but lags in environmental quality compared to northern alternatives"}), and the system logs this revision as a new insight on the timeline. Since Amy's reasoning is externalized and persistently linked to evidence, reflection becomes systematic and retrieval-based.

As exploration continues into Asian cities, Amy encounters a clash between emerging patterns: while exploring Bangkok, her \textit{"cheap but crowded"} storyline conflicts with strong cultural diversity metrics. Rather than overwrite the earlier path, she right-clicks the Bangkok paragraph and selects \hlcb{narrative}{\textit{Create Branch}}, forking two parallel narratives: one on budget cultural immersion and another on premium sustainability. The \hlcb{reflection}{\textit{Insight Timeline}} records the divergence, preserving both branches with their linked evidence so each can develop independently without losing coherence. Narrative branching allows her to externalize and pursue divergent reasoning paths while maintaining traceability to the decisions.

With multiple threads in motion, Amy steps back for strategic oversight. She opens the \hlcb{reflection}{\textit{Inquiry Board}} to review resolved, active and and stalled questions. Jumping from inquiry cards to their linked narrative sections, she abandons weaker paths and focuses on analyses with more substantial evidence, keeping exploration aligned with the evolving story. Inquiry-level reflection allows Amy to monitor the status and progression of her analytical questions, assess which lines of reasoning remain productive, and strategically redirect effort toward more promising directions.

\textbf{\hlcb{intergration}{Integration}}. When several threads feel coherent enough to communicate, Amy switches to \hlcb{intergration}{\textit{Story Mode}}. The system assembles her active claims, branches, and embedded views into a structured draft. The resulting article aligns each claim with supporting evidence and preserves transparency to alternative branches explored during analysis. Story Mode enables narrative synthesis into a cohesive output where every claim remains connected to its supporting visualizations and reasoning context, producing a defensible and transparent account of her analytical process.

\subsection{Implementation Details}\label{sys/details}

Narrative Scaffolding is implemented as a web application with a Python Flask backend and React Next.js frontend. We employ a chain-of-thought~\cite{wei2022chain} prompting strategy to make LLM reasoning more explicit and context-aware. GPT-5 was used to handle higher-level narrative generation tasks that require longer context windows and complex reasoning for storyline development, ensuring faithful mappings between user intent and chart configurations. GPT-4o-mini was used to manage lightweight operations where responsiveness is critical, such as insight extraction, inquiry summarization, and reflective nudges. This approach was chosen based on empirical testing that balanced consistency, speed, and task-specific performance.

Data visualizations utilize Vega-Lite for chart rendering and Leaflet.js for interactive maps. Text-to-SQL functionality leverages VannaAI~\cite{vannaAI} with OpenAI embedding and few-shot training for query consistency. The system is designed to be model-agnostic, allowing for easy substitution of LLMs as capabilities evolve. Complete prompt templates and configuration details are provided in supplementary materials.

\section{Evaluations}

Our evaluation spanned user and system evaluations. We aimed to assess the usability and effectiveness of \texttt{Narrative Scaffolding} (referred to as \texttt{NS} in the following sections), as well as its reliability in handling underspecified narrative input from users. 

To assess whether \texttt{NS} achieves the design goals identified in our formative study (sec ~\ref{sec:formative}), we structured our evaluation around five research questions. Each of the first four questions corresponds to a design goal, examining how the system's features translate into observable benefits during analysis. The fifth question examines system-level reliability, which also relates to \textbf{DG1}'s emphasis on handling vague input.

\begin{enumerate}
    \item[RQ1][\textbf{Workflow Fluency}] How does \texttt{NS} support fluid transitions from vague inquiries to evidence-grounded insights (\textbf{DG1}), and what impact does this have on the integration of exploration and narrative construction?
    
    \item[RQ2][\textbf{Reflection Depth}] How does \texttt{NS} help users externalize their reasoning and maintain continuity of reflection (\textbf{DG2}) as their analysis evolves over time?

    \item[RQ3][\textbf{Exploration Divergence}] How does \texttt{NS} support nonlinear exploration and alternative reasoning (\textbf{DG3}), enabling users to explore broader analytical space and integrate more evidence?
    
    \item[RQ4][\textbf{Reasoning Traceability}] How does \texttt{NS} support traceable reasoning (\textbf{DG4}), making the analytical process behind insights more visible and defensible?
    
    \item[RQ5][\textbf{Output Reliability}] How reliably does \texttt{NS} produce suitable analytical outputs when prompted with vague, ambiguous, or underspecified narratives (\textbf{DG1})?
\end{enumerate}

\noindent{}The first four research questions focus on assessing the impact on users' workflows and output, while the fifth research question investigated the system's capability. The institutional ethics review board approved both studies. 
\begin{figure*}[!tbp]
  \centering
  \begin{minipage}[b]{0.49\textwidth}
    \includegraphics[width=\textwidth]{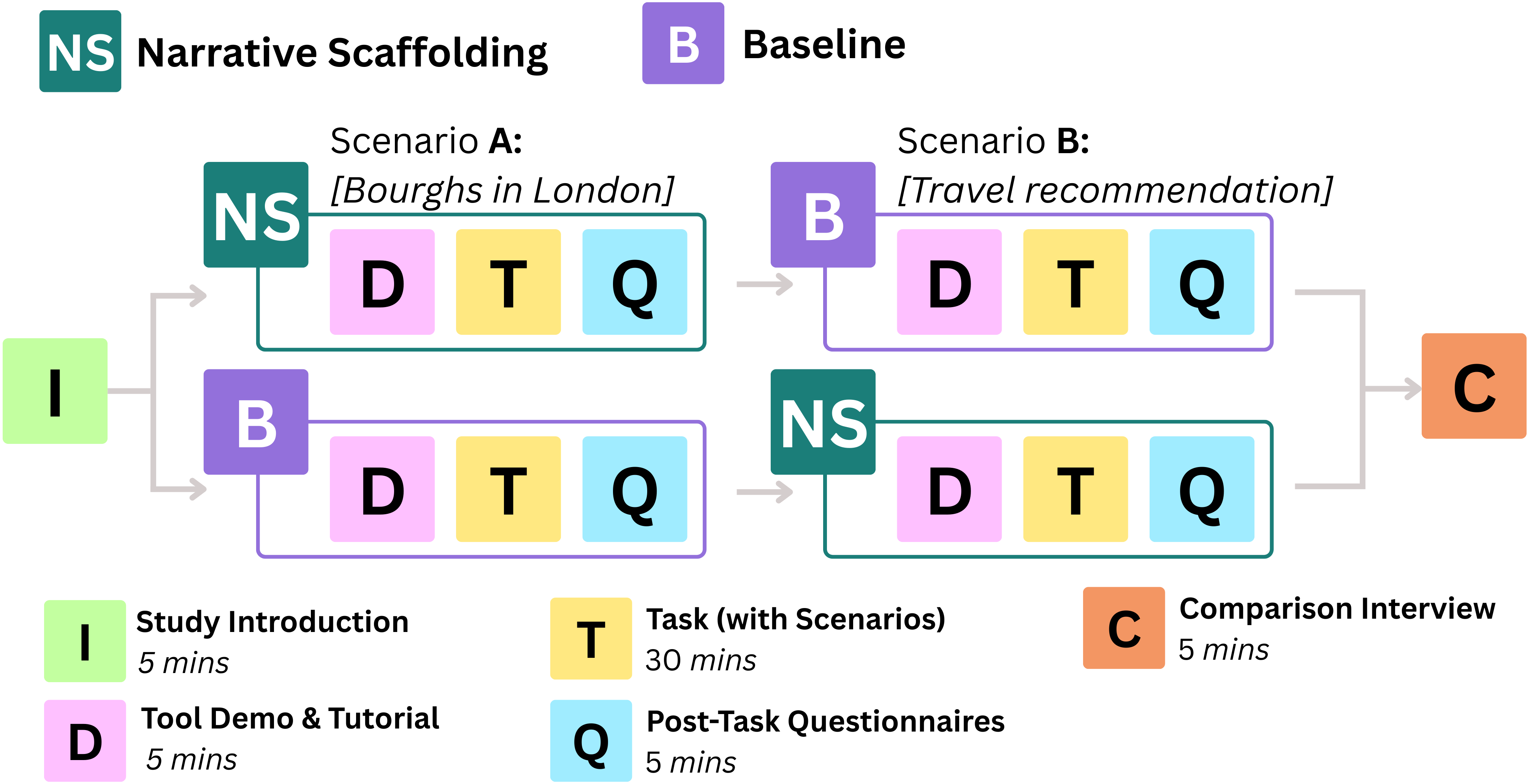}
        \caption{\textbf{Overview of the study procedure.} Participants completed both Narrative Scaffolding (NS) and Baseline (B) conditions in counterbalanced order.}
    \label{fig:procedure}
  \end{minipage}
  \hfill
  \vspace{0.1in}
  \begin{minipage}[b]{0.49\textwidth}
    \includegraphics[width=\textwidth]{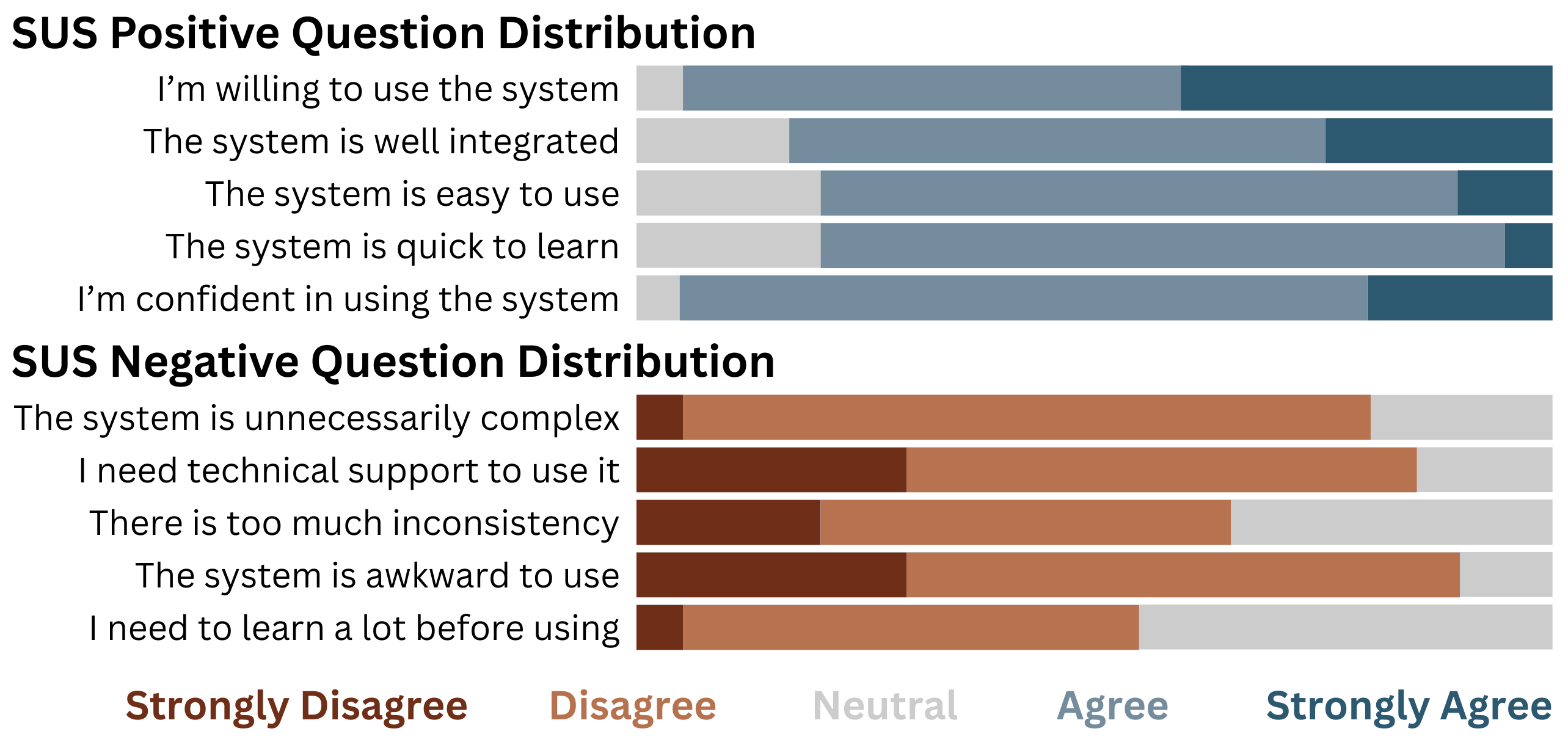}
    \caption{\textbf{SUS questionnaire results.} Positive items (top) received strong agreement, while negative items (bottom) were largely disagreed with.}
    \label{fig:sus}
  \end{minipage}
\end{figure*}

\subsection{Study 1: User Evaluation (RQ1-RQ4)}

We conducted a within-subject evaluation (N=20) to examine whether NS enables more fluid workflows (\textbf{RQ1}), deeper reflection (\textbf{RQ2}), broader exploration (\textbf{RQ3}), and more defensible narratives (\textbf{RQ4}) compared to a baseline analytics paradigm, where participants explored data primarily through views and composed narratives only after analysis.

\subsubsection{\textbf{Baseline and Apparatus}}

In addition to our \texttt{NS} system, we implemented a \texttt{Baseline} for comparison. The \texttt{Baseline} is a web-based analytics workbench that replicates typical Power BI workflows~\cite{microsoftCopilotIntro}, supporting natural-language querying and visualization generation within a traditional dashboard interface. Participants could issue queries, apply filters, generate charts, and export findings using conventional point-and-click interactions supplemented by conversational AI assistance. To ensure a fair comparison focused on workflow design rather than technical capabilities, both systems operated on the same data schema, text-to-SQL translation, and semantic view generation pipeline (\ref{sys:pipeline}). For the final integration stage, we exported tabular data from generated visuals via the Power BI REST API to generate narrative summaries of trends, comparisons, and anomalies~\cite{microsoftPowerBINarrative}. 

This web-based approach allowed participants to join sessions remotely via video conferencing, while also enabling precise logging of user interactions and timing data. The only difference between \texttt{NS} and \texttt{Baseline} lay in the interaction paradigm, isolating effects of interaction structure and reasoning support from differences in analytic functionality.

\subsubsection{\textbf{Method}} We recruited 20 participants (11 male, 9 female; age 23--66, M = 32) through Prolific, using a screening process to ensure prior experience with professional visualization tools. All participants reported using Power BI or Tableau at least once per week. The sample included 7 academic researchers, 7 managers, 5 data analysts, and 1 healthcare professional.

\noindent{}[\textbf{Tasks}]: Participants completed two tasks, each requiring them to write a data narrative recommending options based on multi-factor datasets with eight factors each. Task A involved recommending boroughs for long-term housing value, safety, and community well-being using borough-level data (housing prices, ethnicity, schools, libraries, renting prices, income, crime, gyms, restaurants). Task B involved recommending travel destinations using destination-level data (cost of stay, safety, reviews, diversity, environmental quality, attraction coverage, popularity, crowd dynamics). Both datasets were designed to be balanced in dimensionality and variance to ensure performance differences could be attributed to the systems rather than task complexity. 

\noindent{}[\textbf{Procedure}]: We adopted a within-subject design where each participant used both \texttt{NS} and \texttt{Baseline}, with tool order counterbalanced to mitigate learning effects (Figure ~\ref{fig:procedure}). Sessions began with informed consent, demographics collection, and system introduction. For each system, participants received a tutorial, completed the assigned task while thinking aloud~\cite{eccles2017think} (explicitly verbalizing reflection, exploration, and insights), and completed a post-task survey evaluating usability, workload, and experience. After both systems, participants completed a final comparison survey reflecting on workflow differences. Each session lasted approximately 90 minutes, and participants received \textit{\$20} USD compensation upon completion. Study instructions and survey instruments are provided in the supplementary materials.


\subsubsection{Measures and Analysis}

We collected both perception and interaction data to evaluate the two systems comprehensively. We analyzed post-task survey responses that included 7-point Likert scale questions on workflow efficiency, exploration, reflection, and explainability. To assess usability and workload, we incorporated the SUS~\cite{brooke1996sus} and NASA-TLX~\cite{hart2006nasa}, alongside self-assessments of the clarity and defensibility of participants' final narratives. Interaction logs captured differences in analysis patterns, including query diversity, revisions, and visualizations created, with temporal measures normalized across sessions. In addition, we recorded think-aloud protocols during task sessions to capture participants' reasoning processes in situ. These transcripts were then thematically analyzed to account for the reflection and exploration process. We compared survey responses between the Baseline and NS conditions using the Wilcoxon signed-rank test, given the ordinal nature of Likert-scale data. For continuous temporal measures, we applied Welch's t-test when the normality assumptions were met and the Mann--Whitney U test otherwise. To control for false discovery due to multiple comparisons, we applied the Benjamini--Hochberg procedure~\cite{benjamini1995controlling}. We report Cohen's d effect 
sizes for all parametric comparisons to indicate the magnitude of practical 
differences between conditions.

\subsection{Study 2: System Evaluation (RQ5)}

We evaluated system robustness to vague narrative input, addressing a key requirement for narrative-first interfaces: the ability to translate underspecified user intentions into actionable analytical starting points. 

\subsubsection{\textbf{Evaluation Material}}
To assess system robustness to vague input, we constructed a pool of 100 prompts. We first selected 43 vague prompts collected from the user study that exhibited natural underspecification in areas such as timespan, entity, or aggregation level. We used an LLM to expand the pool by generating additional prompts that followed similar templates but varied in domain and phrasing. To ensure ecological validity, prompts were required to be underspecified in key dimensions (i.e, missing temporal scope, unclear entity reference, or ambiguous level of detail).

\subsubsection{\textbf{Expert Rating}}
We conducted an expert evaluation to assess how well the system outputs addressed vague prompts. Six experts were recruited, with eligibility limited to individuals reporting a minimum of five years of experience in visualization literacy (M=9.5 years). These experts were selected for their ability to evaluate whether visualization choices appropriately match analytical intent and support iterative exploration. Each expert was assigned 50 prompts. Each of the 100 prompts was independently rated by three different experts, resulting in a total of 300 ratings across the pool.

Each visualization was rated on two dimensions using 7-point Likert scales: \textit{(i)} how well the chart plan aligned with the intent of the vague prompt, and \textit{(ii)} the extent to which the chart plan supported further exploration. We averaged the three ratings per visualization to obtain a single score on each dimension.

\section{Results}
We present an analysis of survey responses and participant interactions ($P_1 - P_{20}$) with both conditions. Our results show that \texttt{NS} enabled participants to generate insights more efficiently, engage in deeper and more continuous reflection, explore broader analytical alternatives, and produce more defensible narratives compared to the baseline condition. \texttt{NS} received an average System Usability Score (SUS) of \textit{79.8} (Figure~\ref{fig:sus}), demonstrated high usability (\textit{Excellent} according to \cite{bangor2009determining} and passing the cutoff score, 68, for production by a large margin). $P_{11}$ described the system as feeling "like a reliable companion" during analysis. 

\begin{figure*}[ht]
    \centering
    \includegraphics[width=1.05\linewidth]{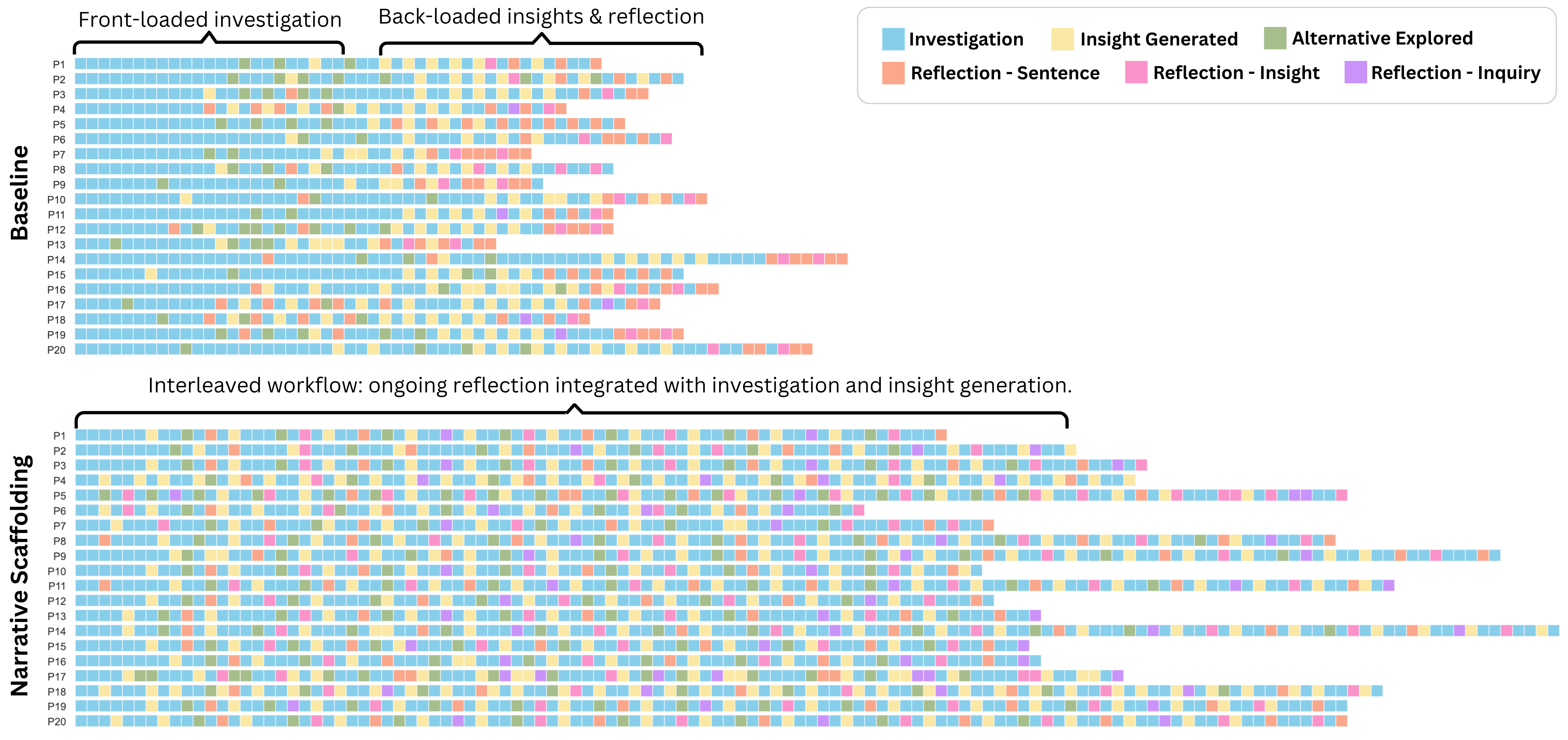}
    \caption{\textbf{Events timeline visualization comparing analytical workflows across Baseline and \texttt{Narrative Scaffolding conditions}.} Each horizontal bar represents one participant's session with color-coded activities. Baseline (top) exhibits a front-loaded investigation pattern with insights and reflection concentrated toward the end of sessions, reflecting a separation between exploration and reasoning articulation. \texttt{Narrative Scaffolding} (bottom) demonstrates an interleaved workflow where investigation, insight generation, and reflection occur continuously throughout sessions, with substantially more alternative exploration and inquiry-level reflection.}
    \label{fig:timeline2-result}
\end{figure*}

Figure~\ref{fig:timeline2-result} provides a timeline view of participant activity across conditions. It highlights how NS sessions were denser and more varied, with frequent reflection and branching events, whereas Baseline sessions consisted largely of investigation alone. This overview illustrates the systemic advantage of NS, which we unpack across workflow fluency (RQ1), reflection (RQ2), exploration divergence (RQ3), and explainability (RQ4).


\subsection{[RQ1]: Workflow Fluency} 
\subsubsection{Participants Generated More Insights Through Deeper Engagement} 
Participants took slightly more time to complete tasks with \texttt{NS} (M = \textit{31.59}, SD = \textit{0.92} min) compared to \texttt{Baseline} (M = \textit{29.84}, SD = \textit{0.87} min). This result is unsurprising, as our goal was to support deeper engagement with data through \texttt{NS} rather than to accelerate task completion. 

However, this similar time investment resulted in substantially greater analytical productivity. \texttt{NS} participants generated significantly more insights (M = \textit{15.65}, SD = \textit{6.95}, p < 0.001) compared to the \texttt{Baseline} (M = \textit{6.85}, SD = \textit{1.69}). This increased insight volume translated into dramatically improved efficiency. The time per insight was significantly lower in \texttt{NS} (M = \textit{2.31} min/insight, SD = \textit{0.76}, p < 0.001) compared to \texttt{Baseline} (M = \textit{4.62} min/insight, SD = \textit{1.22}), representing faster insight articulation. This result aligns with our design goal of enabling more efficient insight generation during exploratory analysis.

\subsubsection{Participants Experienced Reduced Cognitive Load} 
In addition to generating significantly more insights, \texttt{NS} also reduced perceived cognitive load (Figure~\ref{fig:likert-result1}, A). We used NASA-TLX to measure the perceived workload associated with each system. Compared to \texttt{Baseline}, \texttt{NS} had significantly lower mental demand (M = \textit{3.00} < \textit{4.45}, p = \textit{.015}), required significantly less effort (M = \textit{2.95} < \textit{4.60}, p < \textit{.001}), resulted in substantially less frustration (M = \textit{3.10} < \textit{4.75}, p = \textit{.001}), and led to significantly better perceived performance (M = \textit{2.25} < \textit{3.60}, p = \textit{.024}), where lower scores indicate better performance. This suggests that \texttt{NS} made the analytical process feel more natural and manageable despite the increased analytical output.

\subsubsection{Participants Found Direction and Generated Structured Output More Easily} 
\begin{figure*}[ht]
    \centering
    \includegraphics[width=1\linewidth]{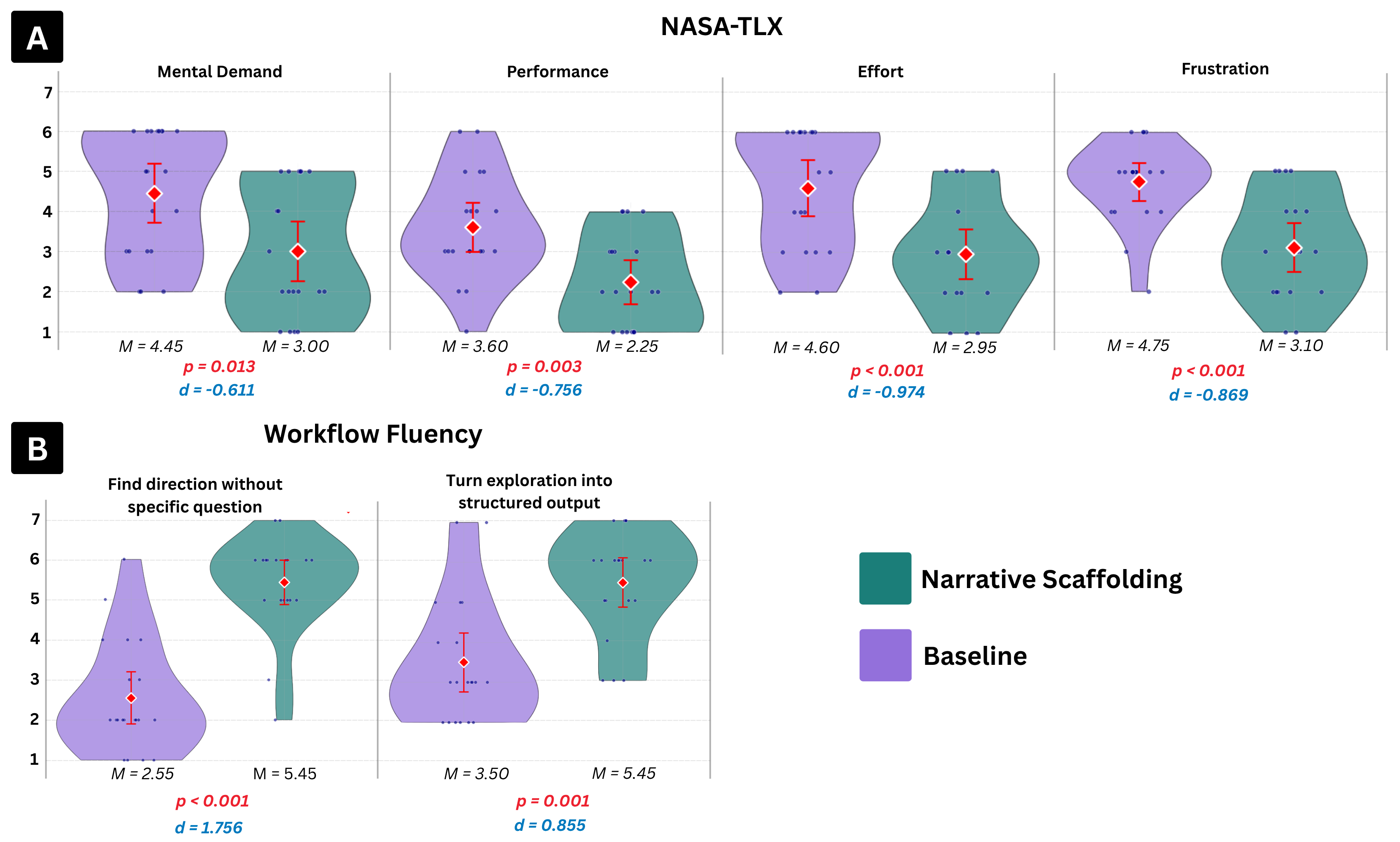}
    \caption{\textbf{Workflow Fluency and Cognitive Load} Comparison of (A) NASA-TLX workload dimensions and (B) workflow fluency measures between NS and Baseline conditions. NS demonstrated significantly lower cognitive demands (p < .05) and improved support for exploration initiation and output structuring. Violin plots show response distributions with means marked by red diamonds.}
    \label{fig:likert-result1}
\end{figure*}

\texttt{NS} participants rated significantly higher ability to find direction without specific questions in mind (Figure~\ref{fig:likert-result1}, B) (\texttt{NS}: M = \textit{5.45}, SD = \textit{0.69}; \texttt{Baseline}: M = \textit{2.55}, SD = \textit{0.93}, p < \textit{.001}). Participants appreciated how \texttt{NS} helped them navigate uncertainty in early exploration. $P_6$ noted the system could \textit{"quickly see what's important and suggest possible directions,"} while $P_{12}$ described how the previews eliminated the challenge of \textit{"starting at abstract ideas trying to figure out where to start."}

\texttt{NS} also significantly improved participants' ability to turn exploration into structured output (Figure~\ref{fig:likert-result1}, B) (\texttt{NS}: M = \textit{5.45}, SD = \textit{0.89}; \texttt{Baseline}: M = \textit{3.50}, SD = \textit{1.06}, p = \textit{.002}). $P_7$ found that "turning exploration into a report feels almost effortless," and $P_{17}$ highlighted how the system "transforms explored branches into a structured story with one click." These ratings align with the behavioral efficiency measures, indicating that participants both experienced and recognized the workflow improvements provided by \texttt{NS}, supporting \textbf{DG1}'s goal of fluid transitions from vague inquiries to evidence-grounded insights.

\subsection{[RQ2]: Reflection Depth}

\begin{figure*}[ht]
    \centering
    \includegraphics[width=0.75\linewidth]{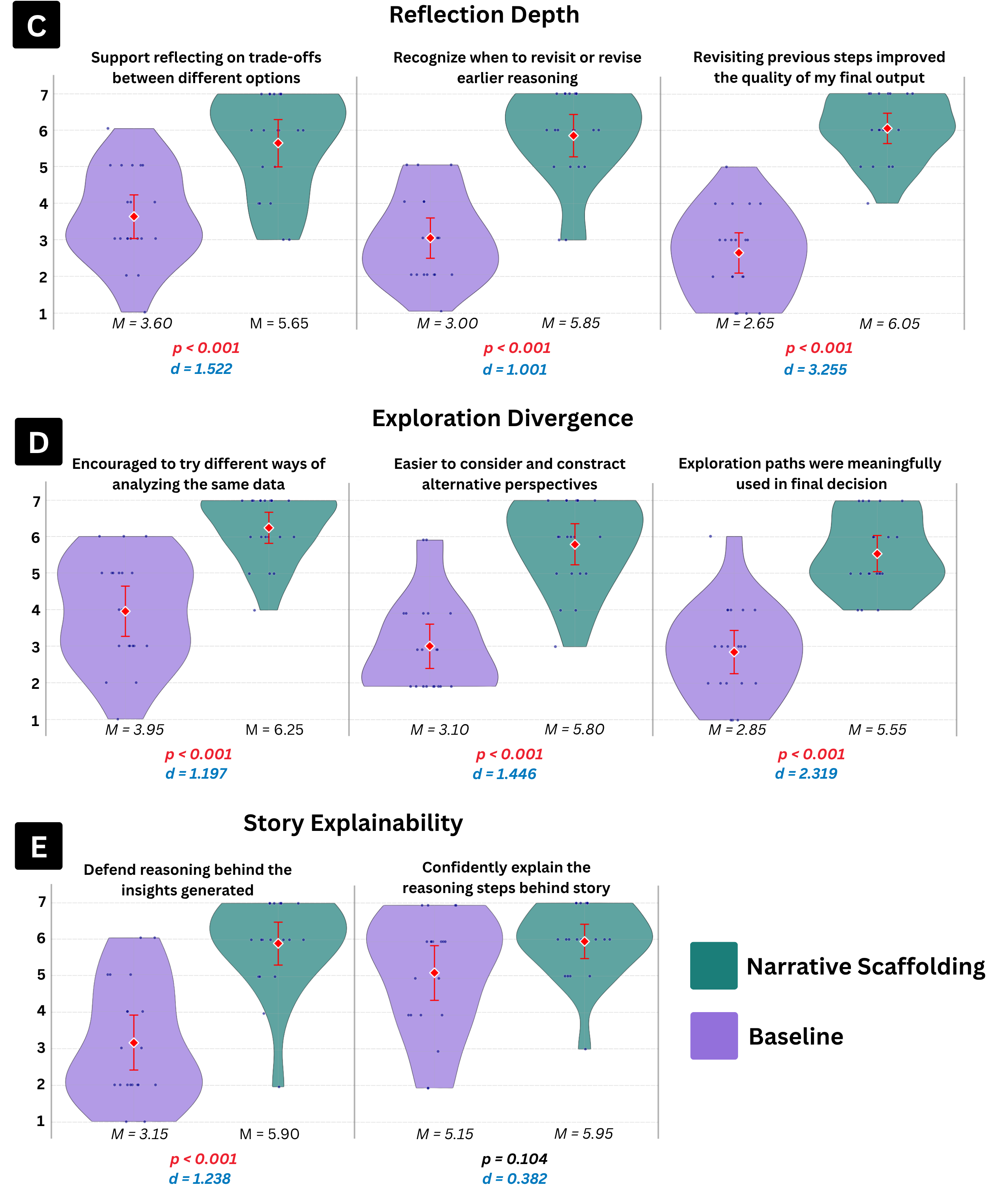}
    \caption{\textbf{Reflection, Divergence, and Explainability} Self-reported measures of (C) reflection depth, (D) exploration divergence, and (E) story explainability across conditions. NS participants reported significantly stronger support (p < .05) for revisiting reasoning, exploring alternatives, and defending conclusions. Violin plots show response distributions with means marked by red diamonds.}
    \label{fig:likert-result2}
\end{figure*}

\subsubsection{Participants Engaged in More Frequent and Deeper Reflection} Across three levels of reflection: sentence-level, insight-level, and inquiry-level, \texttt{NS} participants engaged in significantly more reflections (M = \textit{14.80}, SD = \textit{3.69}, p < 0.001) compared to \texttt{Baseline} (M = \textit{6.50}, SD = \textit{1.91}). The improvement varied substantially by reflection type: sentence-level showed a \textit{1.2×} increase, insight-level increased \textit{3.7×}, and inquiry-level demonstrated a \textit{13.0×} improvement in \texttt{NS} compared to \texttt{Baseline}.

\begin{figure*}[ht]
    \centering
    \includegraphics[width=1\linewidth]{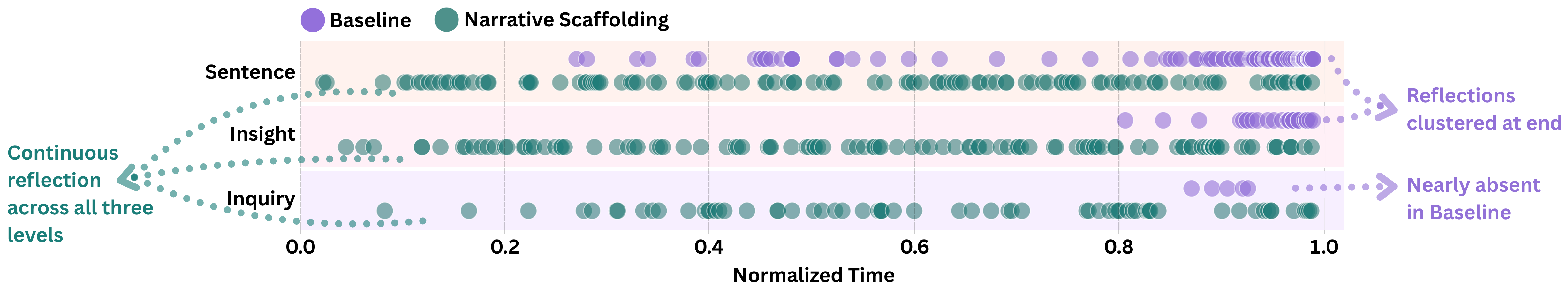}
    \caption{\textbf{Temporal distribution of reflection events across three types throughout the analysis session.} Each dot represents a reflection event by a participant, positioned along normalized session time (0.0 = start, 1.0 = end). Narrative Scaffolding participants engaged in reflection more frequently and distributed more evenly across the session, particularly at the sentence and insight levels. Baseline participants showed concentrated reflection activity toward the end of sessions, with notably fewer inquiry-level reflections overall.}
    \label{fig:reflection-result}
\end{figure*}

The temporal distribution of reflective activity also differed between conditions (Figure~\ref{fig:reflection-result}). \texttt{NS} participants distributed reflection events more evenly across the session timeline, while \texttt{Baseline} participants concentrated most reflection events in the final portions of their sessions. As $P_1$ noted: \textit{"With [NS] it felt natural to pause along the way, look back at what I'd written."}

\subsubsection{Participants Reported Higher Reflection Quality and Usefulness}

While \texttt{NS} generated more reflection events, we examined whether these were actually useful for reasoning improvement. We measured reflection events that were followed within 30 seconds by an edit to the related narratives. \texttt{Baseline} participants showed higher reflection-edit rates (\textit{58.8\%} led to immediate edits) compared to \texttt{NS} participants (\textit{35.2\%}). However, this lower rate does not necessarily indicate a reduction in quality, as not all helpful reflections require immediate edits.

To assess usefulness, we examined self-reported measures of reflection quality (Figure~\ref{fig:likert-result2}, C). \texttt{NS} participants felt significantly better supported in recognizing when to revisit reasoning (\texttt{NS}: M = \textit{5.85}, SD = \textit{1.23}; \texttt{Baseline}: M = \textit{3.00}, SD = \textit{1.17}, p < \textit{.001}) and reflecting on trade-offs between options (\texttt{NS}: M = \textit{5.65}, SD = \textit{1.39}; \texttt{Baseline}: M = \textit{3.60}, SD = \textit{1.27}, p < \textit{.001}). Most importantly, participants reported that revisiting previous steps significantly improved their final output quality (\texttt{NS}: M = \textit{6.05}, SD = \textit{0.89}; \texttt{Baseline}: M = \textit{2.65}, SD = \textit{1.18}, p < \textit{.001}).

\texttt{NS} participants described using reflection for ongoing reasoning validation rather than just problem-fixing. $P_{13}$ noted relying on \textit{"the timeline to check when and why earlier reasoning might need to be revisited,"} while $P_{10}$ found \textit{"the inquiry board helps me track which questions still need attention."} These findings indicate many reflection events in \texttt{NS} serve to confirm conclusions or validate reasoning. Despite lower reflection-edit rates, \texttt{NS} participants reported significantly higher perceived benefits for reasoning quality, suggesting that NS successfully externalized inquiry intent and maintained reflective continuity (\textbf{DG2}).

\subsection{[RQ3]: Exploration Divergence}

\subsubsection{Participants Explored More Alternatives and Factors}

With \texttt{NS}, participants considered significantly more alternatives during their analysis (M = 8.75, SD = 3.03, p < 0.001) compared to \texttt{Baseline} participants (M = 3.25, SD = 1.58). They also explored a greater breadth of available factors, examining on average 7.0 out of 8 possible factors (SD = 0.9), compared to Baseline participants who explored 5.3 factors (SD = 1.4). As $P_8$ explained, \textit{"With NS I felt like I could easily try different takes on the data, instead of sticking to one angle."}

Self-reported measures confirmed that participants felt \texttt{NS} better supported exploratory breadth (Figure~\ref{fig:likert-result2}, D). Participants rated \texttt{NS} significantly higher for encouraging different alternatives for analyzing the same data (\texttt{NS}: M = 6.25, SD = 0.91; \texttt{Baseline}: M = 3.95, SD = 1.47, p < .001) and for making it easier to consider and contrast alternative perspectives (\texttt{NS}: M = 5.80, SD = 1.20; \texttt{Baseline}: M = 3.10, SD = 1.29, p < .001). Participants ($P_4, P_7, P_{13}, P_{16}, P_{17}$) themselves described this sense of breadth, noting that \textit{"[Baseline] made me tunnel on one or two things, but [with NS] I actually considered a range of options before narrowing down."} -- $P_4$

$P_{15}$ emphasized how NS encouraged breadth across factors: \textit{"The system had nudged me to look at other dimensions when I reflected, like community well-being alongside cost."} Together, these findings suggest that NS not only increased the number of alternatives participants explored but also helped them consider broader coverage of available factors, creating more opportunities to surface contrasting perspectives.

This broader and more varied exploration pattern is also visible in the activity timelines (Figure~\ref{fig:timeline2-result}): NS participants interleaved investigation with reflection and branching throughout their sessions, while Baseline activity remained concentrated on linear investigation. This illustrates how NS not only encouraged participants to try more alternatives but also supported a workflow where reflection and exploration reinforced one another.

\subsubsection{Participants Integrated More Evidence into Decision-Making}

Beyond exploring more alternatives, \texttt{NS} participants integrated more of their exploration into final decisions. They incorporated significantly more factors into their decision-making process (M = \textit{3.8}, SD = \textit{0.7}) compared to \texttt{Baseline} participants (M = \textit{2.5}, SD = \textit{0.7}). This suggests that \texttt{NS} participants not only explored more broadly but also utilized their exploration more meaningfully. Participants emphasized that the more exhaustive exploration in \texttt{NS} translated into having more actionable options to weigh. As $P_{11}$ explained, \textit{"Because I had looked at more angles, I actually had choices, I could go in a few different directions instead of just one."} Similarly, $P_6$ elaborated, \textit{"Having more alternatives in [NS] meant I didn't have to stick with my first idea [...] I could line up different options and actually remember how they compared, which made it easier to choose from."}

Self-reports reinforced this pattern (Figure~\ref{fig:likert-result2}, E), participants rated \texttt{NS} significantly higher for ensuring exploration paths were meaningfully used in their final story (\texttt{NS}: M = \textit{5.55}, SD = \textit{1.05}; \texttt{Baseline}: M = \textit{2.85}, SD = \textit{1.27}, p < \textit{.001}). These results indicate that \texttt{NS} enabled not only wider exploration but also more systematic integration of diverse evidence into coherent analytical conclusions (\textbf{DG3}), providing participants with more usable options when making decisions.

\subsection{[RQ4]: Reasoning Traceability}

We assessed whether NS improved users' ability to trace and justify their analytical reasoning process. Participants using \texttt{NS} reported significantly greater confidence in defending the reasoning behind their insights (M = \textit{5.90}, SD = \textit{1.25}) compared to \texttt{Baseline} (M = \textit{3.15}, SD = \textit{1.60}, p < \textit{.001}). Participants described how timelines and provenance features supported justification. $P_2$ explained, \textit{"The timelines show the whole process which guarantees transparency and provides relevant evidence."} $P_{13}$ noted, \textit{"Since the insights were linked to evidence and captured in the timeline, it is pretty straightforward to justify my reasoning."} Similarly, $P_{17}$ emphasized, \textit{"Since every captured insight is automatically linked back to its supporting data and visualizations, I would always have clear evidence to justify my claims."} $P_{20}$ summarized this benefit as, \textit{"I felt more confident that I could defend my choices because I had pulled in evidence from different places, not just one metric."}

When asked about presenting conclusions to others, participants also reported higher average confidence with \texttt{NS} (M = \textit{5.95}, SD = \textit{1.00}) than with \texttt{Baseline} (M = \textit{5.15}, SD = \textit{1.60}). However, this difference was not statistically significant (p = \textit{.112}).  Participants ($N = 13$) rated themselves as confident presenting with \texttt{NS}, citing features such as organized evidence, visible branches, and insight timelines, which serve as valuable sources of support. For instance, $P_1$ stated, \textit{"With all evidence organized, branches visible, and a polished story generated, I'd feel confident explaining my analysis to anyone."} Similarly, $P_{10}$ noted, \textit{"Every insight is backed by reasoning, and the story mode is also helpful; therefore, it will aid in justifying my recommendations."}

Among these same participants, ($N = 9$) also rated themselves as confident presenting with \texttt{Baseline}. In these cases, confidence stemmed from their own skills rather than the tool itself: $P_1$ remarked, \textit{"I feel really confident because once I make a presentation, I will be able to address the weak points in my insights and conclusions."} Similarly, $P_{12}$ explained, \textit{"I'd be comfortable with the recommendations themselves, but less confident explaining the full path I took to get there since some of that context lives outside the tool."}

This contrast suggests that while participants often considered themselves capable of presenting regardless of the tool, the source of their confidence differed. In \texttt{NS}, confidence was rooted in system-supported provenance and narrative structure (\textbf{DG4}), whereas in \texttt{Baseline}, confidence was self-reliant, grounded in personal presentation skills rather than the analytic workflow. 

\subsection{[RQ5]: Output Reliability}

 We analyzed expert ratings ($E_1 - E_6$) along two dimensions: (i) alignment with the intent of vague prompts and (ii) support for further exploration. The first captures whether outputs were judged as plausible and relevant responses to the expressed inquiry, recognizing that vague prompts can admit multiple reasonable interpretations. The second captures whether outputs were seen as useful entry points for continuing analysis, since vague prompts are rarely end goals but instead serve to open inquiry. For each dimension, we report quantitative results followed by qualitative feedback from experts on cases they felt did not work well.

Expert ratings indicated that the system outputs were generally robust across vague prompts. For intent alignment, 77\% received "agree" or "strongly agree" ratings, 20\% fell between "slightly agree" and "neutral," and only 3\% were rated "slightly disagree" or lower. For supporting exploration, 79\% were rated "agree" or "strongly agree," 18\% between "slightly agree" and "neutral," and 3\% "slightly disagree" or below. These distributions suggest that the large majority of outputs were considered both relevant to the prompt and helpful in sustaining inquiry.

Experts explained that lower ratings often reflected a failure to capture implied intent. While generated dashboards were usually relevant to the surface wording of the prompt, they sometimes missed the deeper analytical angle that vague phrasing implied. For example, $E_3$ noted a prompt about "some neighbourhoods are unexpectedly safe" suggested an outlier-focused analysis, but the system produced generic averages and correlations. Similarly, $E_5$ explained that when asked about \textit{"travel growth differences between countries"} the system produced growth rate visualizations that \textit{"It showed similar percentages across countries, but missed that 10\% growth could mean millions in one place and only thousands in another."} $E_2$ described such cases as \textit{"technically correct but superficial"}. This implies that while the majority of outputs were judged as good matches to the prompts, a smaller subset still fell short by failing to respond to the implied meaning behind vague inquiries. 

In general, experts appreciated that the system consistently produced plausible and relevant outputs, though they also noted occasional reliance on generic templates or literal matches that did not surface distinctive analytical directions. These suggest that the system is sufficient for producing semantically aligned dashboards from vague prompts at scale. Targeted refinements (diversifying templates, capturing implied intent) could further enhance its utility.

\section{Discussion}

\subsection{The Narrative-First Paradigm: Supporting Natural Reasoning Through Writing}

Our findings reveal how narrative construction can function as a primary reasoning interface during data analysis, aligning with narrative psychology's view that storytelling structures how people make sense of complex information~\cite{murray2003narrative, bruner1991narrative}. Rather than treating writing as post-analysis documentation~\cite{zhao2021chartstory, lee2015more}, participants used \texttt{NS} not merely to document completed insights, but as a thinking medium, a space where vague hunches evolved into precise, evidence-backed claims through the act of writing. 

A key aspect of this narrative-as-thinking-medium approach was adapted to how participants actually think. This paradigm accommodated diverse analytical entry points and supported continuous interpretive refinement throughout exploration. Unlike visualization-first tools that require users to translate thoughts into 
structured queries~\cite{setlur2016eviza, gao2015datatone}, \texttt{NS} 
accommodated diverse entry points---whether specific hypotheses or exploratory 
questions. This addresses the \textit{"gulf of envisioning"}~\cite{subramonyam2024bridging}: the difficulty of translating vague intentions into system-actionable inputs. Participants could write underspecified hunches, generate visualizations, then 
iteratively refine their narrative to pursue new directions---each refinement 
triggering updated views without breaking flow. This fluidity reduced friction not just during early-stage exploration~\cite{wongsuphasawat2019goals, kandel2012enterprise}, but throughout the entire sensemaking process, as participants' understanding evolved and required new framings.

Beyond these cognitive benefits for users, narrative-first design provided unique system-level benefits by making reasoning evolution visible and trackable. Externalizing thoughts in narrative form allowed participants to monitor how their assumptions shifted and revisit earlier claims. This highlights how narrative construction serves dual purposes: supporting human reasoning while providing systems with insight-level data that traditional interaction logs cannot capture~\cite{ragan2015characterizing, north2011analytic}.

Our results demonstrate that when narrative construction becomes the primary reasoning interface, systems can scaffold the process of thinking rather than just its outputs. This opens possibilities for more sophisticated provenance, adaptive scaffolding, and reasoning-aware recommendations. However, some participants noted that highly quantitative comparisons --- such as spatial relationships (e.g., visually scanning which of 50 cities cluster together geographically) or multi-dimensional trade-offs (e.g.,simultaneously meeting cost and safety thresholds) --- were difficult to express through text alone, as these tasks benefit from parallel visual processing. These limitations suggest that while narrative-first design offers powerful scaffolds for reasoning, future mixed-initiative systems should also integrate complementary modalities that better support quantitative and spatial thinking.

\subsection{Redirecting Cognitive Effort and Supporting Productive Friction}

Most interactive systems aim to reduce user effort under the assumption that easier interactions lead to better outcomes. However, cognitive load theory distinguishes between productive cognitive effort that supports learning and extraneous overhead that impedes it~\cite{kalyuga2011cognitive, sweller1994cognitive}. Rather than simply minimizing effort, our findings suggest the importance of redirecting cognitive resources toward valuable reasoning activities. Our findings highlight that participants using \texttt{NS} reported significantly lower mental demand, effort, and frustration while engaging in more complex reasoning. For example, they generated 2.3 times more insights and conducting 2.7 times more reflection events. Also, they engaged in reflection throughout the whole process instead of only towards the end. This pattern suggests that \texttt{NS} reshaped how participants allocated cognitive resources during analysis.

These results suggest a nuanced view of the relationship between analysis depth and cognitive burden. Our findings reveal two distinct types of cognitive load in analytical work. \texttt{NS} reduced overhead load by eliminating cognitive friction around procedural tasks. As $P_{12}$ noted: \textit{"I didn't have to remember where I wrote things down or try to piece together my thinking."} Simultaneously, \texttt{NS} redirected effort toward reasoning activities (reflection, comparison, and synthesis) rather than procedural management. $P_{10}$ described this shift: \textit{"Instead of spending time on how to organize my thoughts, I now focus on what the data actually telling me."} This redistribution of cognitive effort explains why participants could engage in more sophisticated reasoning while experiencing less subjective difficulty.  The system's workflow effectively managed the demands of tracking and reconnecting insights, allowing participants to focus on \textit{interpretive reasoning}.

Overall, our study highlights the importance of optimizing the type of effort users expend rather than simply minimizing user effort, suggesting that design evaluation should focus on whether systems enhance reasoning-focused effort, rather than prioritizing workload reduction solely.

\subsection{Design Implications for Mixed-Initiative Analytical Systems}

A key challenge in designing mixed-initiative analytical systems is determining when and how AI should participate in users' reasoning processes. Our findings reveal that effective analysis requires treating exploration, reflection, and articulation as an integrated process. We identified three interdependent stages: (1) divergence, which creates raw material by generating multiple alternatives and factors; (2) reflection, which evaluates and refines these alternatives; and (3) integration, which incorporates them into defensible conclusions.

Our findings suggest these stages are interdependent. Divergence without reflection produces scattered exploration; reflection without alternatives limits reasoning depth; integration depends on both. This interdependence suggests that individual features, while valuable, may provide incomplete support for deep reasoning. Our findings indicate that sophisticated analysis benefits from systemic design where exploration, reflection, and integration are scaffolded as interconnected processes rather than isolated capabilities.
\subsubsection*{Supporting Divergence Through Alternative Engagement}
Effective divergence depends on helping users engage with multiple alternatives, rather than merely producing superficial options~\cite{suh2024luminate, pu2025ideasynth}. Our bidirectional narrative--evidence coupling illustrates this. Features like \textit{Show View} and \textit{Capture} let users test a narrative hunch with visualizations, then save what they discover back into their reasoning. This cycle helps alternatives develop enough depth to support later reflection. While not every system needs a narrative-first interface, the principle applies broadly: mixed-initiative systems can support divergence by enabling exploration of multiple paths that help develop competing explanations from initial hunches into well-reasoned, evidence-backed alternatives.
\subsubsection*{Provenance Needs to Capture Why, Not Just What}
Productive reflection requires access to the underlying reasoning, not just a record of interactions. Our insight timeline and branching features illustrate one approach: when users create branches or mark shifts, the system captures both the change and its surrounding narrative context, making earlier reasoning states revisitable and comparable. Whether using narratives or other forms, the aim is to make the evolution of reasoning both visible and revisitable. By capturing the reasons behind users' changes in direction, rather than merely tracking their clicks, these systems foster genuine reflection instead of simple interaction replay.

\subsection{Supporting Reasoning Evolution Across Domains}

The principles underlying \texttt{NS} address a challenge that extends beyond data analysis: how to support reasoning that develops over time. A clear example is intelligence analysis, where analysts must synthesize fragmented information into coherent assessments while remaining open to contradictory evidence. \texttt{NS} suggests potential to track reasoning from initial hypotheses to evidence integration and final judgments, helping analysts maintain clear records of their reasoning and remain open to alternative explanations on ambiguous problems~\cite{pherson2019structured, heuer1999psychology}. Similar needs arise in policy development, where decision-makers must manage unresolved questions, weigh competing factors, and show how recommendations evolve from evidence~\cite{hansen2010politics}. In research contexts, tools could likewise benefit from insight timelines that track the progression of hypotheses across literature reviews, experiments, and analysis.

More broadly, these examples highlight a shared design gap: most platforms capture either final outputs (documents, decisions) or low-level interactions (clicks, searches) but miss the interpretive work that connects them. By treating reasoning threads as first-class objects, systems could help users revisit earlier assumptions, compare alternative explanations, and refine conclusions over time. Whether in intelligence, policy, or research, our framework highlights opportunities for system architectures that make reasoning evolution explicit, which may improve transparency, accountability, and coherence across domains.

\subsection{Ethical Implications for Analytical Integrity}

AI-assisted analysis systems raise concerns about analytical integrity when algorithmic decisions become invisible to users~\cite{ha2024guided}. We discuss three areas where \texttt{NS}'s design could unintentionally affect reasoning quality: algorithmic bias in view generation, narrative framing effects, and system steering through workflow design.

\subsubsection*{Algorithmic Bias in Semantic Alignment.}

While utilizing LLMs for semantic alignment enables efficient translation from narrative to visualizations, the model generation could contain inherent biases that negatively impact the quality of analysis. Studies have shown that LLMs could exhibit systematic errors in query generation, inconsistent interpretation and visualization selection. When a user writes "Asia is cheaper but more crowded," the system makes multiple interpretive choices: which columns represent "cheap," what aggregation level defines regions, and which chart types show comparisons. These encoded assumptions can privilege certain interpretations while obscuring others. To mitigate these risks, NS maintains human control over AI-generated outputs. Views are generated on explicit user request rather than automatically, and users decide whether to include them in their narrative. The bidirectional capture mechanism provides additional context to prompt context-aware generation.  This human-in-the-loop approach ensures that AI assists rather than dictates analytical reasoning. However, significant risks remain. Users may not recognize when generated queries misrepresent their intent or when visualizations appear plausible but encode problematic assumptions. Future interventions can further mitigate these risks by exposing generated SQL queries for user verification, providing confidence scores for query correctness, and explaining the reasoning behind visualization choices to help users recognize when outputs misrepresent their analytical intent.

\subsubsection*{Narrative Framing Effects.} 

Confirmation bias is a risk that analysts actively work to avoid in rigorous analysis. NS explicitly surfaces this risk by enabling users to start exploration with narrative hypotheses. When users write claims and the system generates aligned views, they may selectively interpret patterns as confirming their narrative while overlooking contradictions. Following design guidelines that emphasize exploring alternatives to reduce confirmation bias, NS supports narrative branching to explore alternative explanations with proactive reflection for revisit. The inquiry board maintains visibility of unresolved questions, encouraging users to examine competing interpretations. However, these mechanisms rely on users recognizing when their reasoning has narrowed and taking action to branch or reflect. Users deeply committed to their initial narrative may not recognize the need for alternatives. Future systems could explore proactive interventions that detect when narrative commitment narrows exploration and suggest contradictory evidence or alternative framings to maintain analytical rigor.

\subsubsection*{System Steering Through Workflow Design.}

While \texttt{NS} reduces cognitive overhead, the workflow design creates structural pressures. Prior work identifies steering challenges in AI-assisted analysis and proposes task decomposition to address them~\cite{kazemitabaar2024improving}. Narrative-driven exploration faces similar challenges at the workflow level and introduces friction through explicit actions for generating views, capturing insights, and creating branches. However, this friction occurs at execution points rather than validation points. Users must independently recognize when alternatives are needed or when reasoning should be questioned, and the system's editorial choices during synthesis remain implicit. While NS makes branching and reflection visible workflow components, these depend on users initiating them. Future systems could explore friction-induced interventions that prompt alternative exploration when analytical patterns suggest narrowing, or make the system's synthesis criteria more transparent.







\section{Limitations and Future Work}

\subsection{Open-Ended Exploration Without Ground-Truth Validation.}  
Our evaluation focused on interpretive sensemaking in open-ended exploratory contexts. We selected tasks with multiple valid interpretations where correctness is context-dependent, and assessed process quality rather than outcome correctness. The datasets have no ground-truth answers, as different stakeholders might legitimately prioritize different factors. 

This design choice means our findings may not generalize to high-stakes analytical contexts where correctness and bias have serious consequences~\cite{wall2017warning, confirmation2025}. In domains such as medical diagnosis~\cite{saposnik2016cognitive} or policy evaluation~\cite{korteling2023cognitive}, reducing confirmation bias and ensuring analytical rigor are critical. While our study suggests \texttt{NS} supports broader exploration and deeper reflection, we cannot claim these process improvements translate into reduced bias or increased accuracy without targeted evaluation in such contexts. 

The limitation extends beyond measurement to the appropriateness of design. High-stakes domains may require scaffolding mechanisms different from those we implemented. For instance, medical diagnosis benefits from structured checklists that enforce consideration of differential diagnoses, while intelligence analysis requires explicit documentation of assumptions and confidence levels. Our narrative-first approach prioritizes flexible exploration and interpretive freedom, which may be less appropriate when analytical protocols must be followed rigorously or when premature narrative commitment could reinforce dangerous biases. 

While \texttt{NS} incorporates design principles aimed at mitigating cognitive bias, their effectiveness in reducing analytical errors remains unverified. Future research should extend this work to high-stakes domains with ground-truth solutions to empirically assess where and how analytical errors occur during narrative-driven exploration. This includes developing methods to detect when reasoning diverges from accuracy, designing interventions that interrupt problematic patterns (e.g., premature commitment to flawed narratives), and understanding optimal timing and modality for such interventions. By studying how narrative scaffolding performs when correctness can be verified, we can identify specific failure modes and develop safeguards for contexts where analytical errors have serious consequences.

\vspace{-1em}
\subsection{Supporting Convergence in Divergent Exploration.} \texttt{NS} creates scaffolds for divergence through branching and alternative exploration but relies heavily on user-driven reflection to synthesize and converge insights. Without active reflection, participants may end up with scattered and unproductive exploration. Our analysis supports convergence through reflection was necessary for productive outcomes. However, systems could better support convergence processes through automated synthesis suggestions or propagation mechanisms that maintain coherence across alternative reasoning paths as users refine and integrate scattered explorations~\cite{suh2025storyensemble}. This could include developing mechanisms to detect when users have sufficient alternatives for meaningful comparison and suggesting synthesis strategies.

\section{Conclusion}

Narrative-driven exploration represents a natural mode of human reasoning with data, yet it carries inherent risks of premature commitment and confirmation bias. Rather than avoiding this approach, we argue the path forward lies in developing AI systems that support and safeguard how people naturally think with data. In this work, we introduced the \texttt{Narrative Scaffolding} framework to address persistent challenges in data-driven sensemaking, particularly in supporting workflow fluency, reflection, divergence, and explainability. Building on the view that data exploration is a process of interpretive reasoning, \texttt{Narrative Scaffolding} treats narrative construction not as an after-the-fact report but as the primary interface for reasoning, coupling writing with semantically aligned view generation, provenance tracking, and branching mechanisms. Our formative studies revealed four key challenges in managing evolving reasoning, and our evaluations of the \texttt{Narrative Scaffolding} system that instantiates the framework showed that it enables more fluid workflows, deeper reflection, broader exploration of alternatives, and more defensible narratives. This work opens new ground for AI-integrated data analysis by treating writing as a reasoning interface for narrative-driven sensemaking. We surface key challenges in balancing human initiative with system guidance and managing trade-offs between exploratory breadth and analytical precision, providing a foundation for future systems that enhance rather than suppress human reasoning with data.

\section*{Usage of Generative AI Statement}
We disclose all instances of GenAI use throughout this research:

\begin{itemize}
    \item \textbf{Prototype development:} Claude-4 was used to assist with coding support during prototype implementation. All generated code was reviewed, modified, and tested by the authors to ensure correctness. 
    
    \item \textbf{Data analysis:} Claude-4 was used to assist with Python notebooks and D3 charts for data analysis. The authors reviewed, validated, and interpreted all analysis results independently. 
    
    \item \textbf{Manuscript writing:} GPT-5 and Claude-4 were used for grammar refinement, sentence restructuring, and improving the clarity and flow of the manuscript. Assistance was limited to language editing and organization. 
\end{itemize}


\bibliographystyle{ACM-Reference-Format}
\bibliography{main}

\appendix
\clearpage
\onecolumn
\section{Appendix}

\subsection{Semantic-Aligned View Generation Pipeline}\label{pipeline}

\begin{center}
\begin{minipage}{0.9\textwidth}
    \centering
    \includegraphics[width=\linewidth]{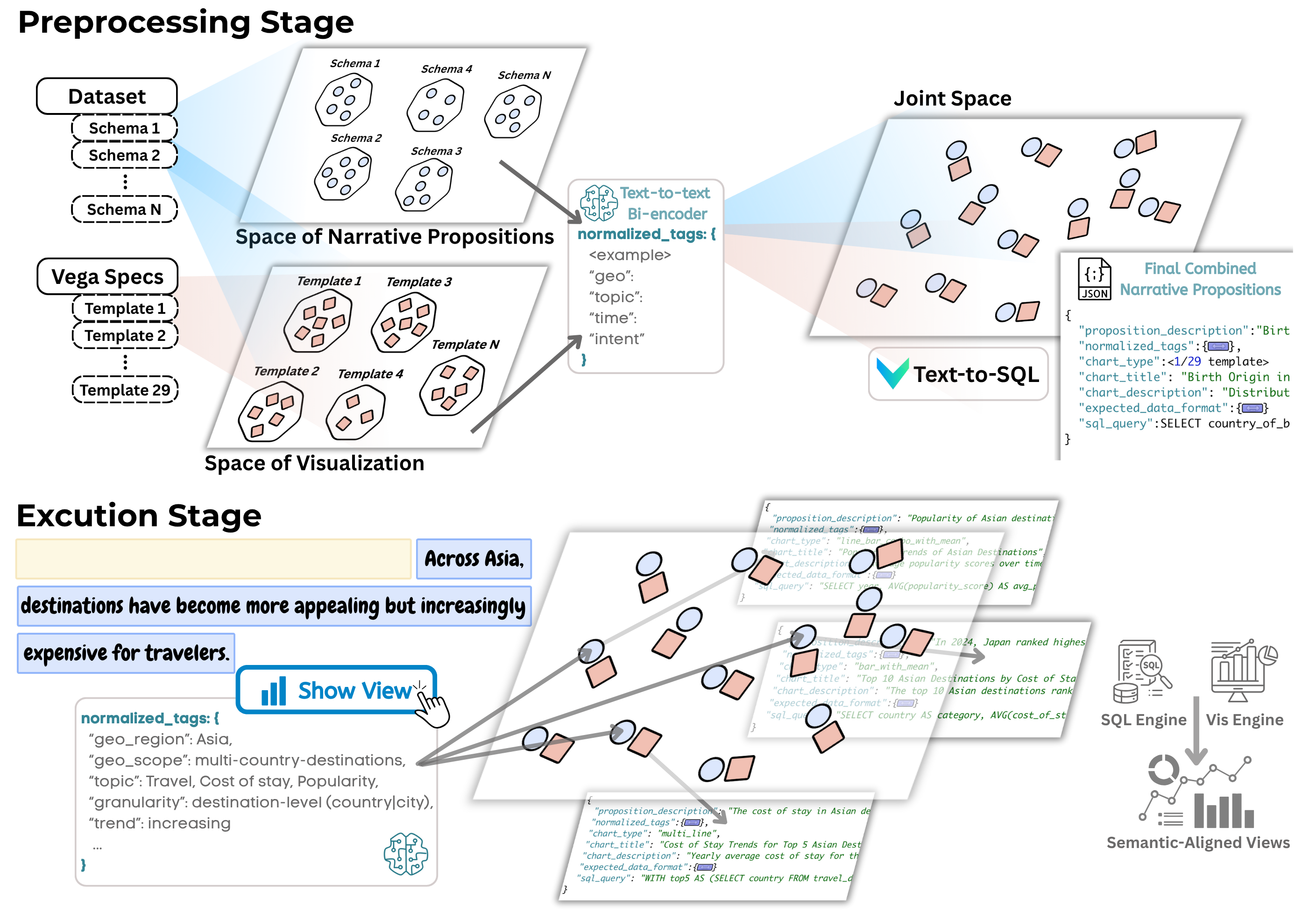}
    \captionof{figure}{\textbf{Semantic-aligned view generation pipeline.} During preprocessing (top), the system generates schema-driven narrative proposition templates and pairs them with visualization specifications. These are normalized into shared semantic tags and embedded into a joint space using a bi-encoder architecture. During execution (bottom), user narratives are normalized into tags and matched against this joint space to retrieve relevant propositions. The system executes corresponding SQL queries, and renders aligned views.}
    \label{fig:pipeline}
\end{minipage}
\end{center}

The narrative proposition space consists of structured templates grounded in the dataset schema, covering common analytical patterns such as ranking, change, composition, outlier detection, per-capita measures, and correlations. For example, from a crime dataset, the system may generate a template such as "In \{\texttt{{year}}\}, \{\texttt{{borough}}\} had the highest \{\texttt{metric}\} at \{\texttt{{value}}\}." Propositions are filtered with constraints (i.e. \texttt{top-k limits}, \texttt{minimum timepoints}) to ensure visibility. For statistical accuracy of the narrative propositions, templates are then instantiated with actual values through schema-driven SQL queries. Constraints, such as requiring multiple time points, further ensure that only meaningful propositions are retained.

The visualization proposition space is derived in parallel, adapting a library of Vega specifications to the dataset's schema. Each chart template defines its expected data format (i.e, scatterplots with \{{\texttt{{x, y, group}}\} or bar charts with \{\texttt{{labels, values}}\}) and is paired with an LLM-generated title and descriptions, ensuring interpretability across various views.

Narrative propositions and visualization descriptions are then normalized into shared tags (i.e \texttt{geo=}, \texttt{topic=}, \texttt{time=}, and \texttt{intent=}) and embedded into a joint semantic space using a bi-encoder architecture~\cite {wang2023improving}. The joint space consolidates all components needed for alignment. It contains the narrative proposition templates, LLM-generated visualization captions, normalized tags, the expected data formats from each visualization specification, and pre-stored SQL queries produced by a text-to-SQL model. By combining narrative and visualization in one embedding space, the system maintains semantic consistency across both directions of investigation.

When users write a narrative and request \textit{\hlcb{investigation}{Show view}}, the system normalizes the text into tags (Figure~\ref{fig:pipeline}, bottom). It matches against the list of narrative propositions. Based on the match, it determines whether a single chart or a group of charts provides the most appropriate alignment with the claim. The corresponding SQL queries are then executed, and the resulting data are rendered into either individual chart templates or grouped into dashboards when multiple views are needed to represent the claim.

\subsection{Prompt Used for Data Story}
\begin{lstlisting}[basicstyle=\ttfamily\small,
                   breaklines=true,
                   frame=single,
                   backgroundcolor=\color{gray!10},
                   numbers=none]
You are a professional data journalist. You will receive an exploration_path as JSON (array of objects). Each object includes at least:

######### INPUT #########

sentence_id
sentence_content
drift_type
other optional fields may be present
######### TASK #########
Produce data_story as a JSON array of 8--15 objects. Each object must contain:

"data_story_sentence": one concise, professional, data-driven sentence suitable for publication.
"ref_id": a single sentence_id or an array of sentence_ids from the exploration_path that support this sentence.
######### GUIDELINES #########

Summarize evidence-driven insights; do not describe analysis steps or methods.
Use objective, newsroom style (e.g., "The data shows," "Analysis reveals," "Destinations such as X and Y emerge").
Cover the main branches of the exploration and converge on key conclusions.
You may condense multiple related exploration sentences into one story point; if so, include all supporting sentence_ids in ref_id (use an array).
Every ref_id must exist in the exploration_path.
Avoid self-referential language (no "I," "we," or "the model").
Do not invent facts; rely only on the content of exploration_path.
Keep each sentence concise and fact-focused.
######### ORDER AND SCOPE #########

Sequence the data_story to follow a logical narrative from overview to key findings to conclusions.
Ensure the story spans the breadth of the exploration_path.
######### STYLE #########

Professional, objective, and data-driven tone.
Present tense unless the data explicitly reference past periods.
Avoid speculation; qualify claims only when supported by the data.
######### OUTPUT FORMAT #########

Output only a valid JSON array.
Do not include any text before or after the JSON.
Each element must include exactly "data_story_sentence" and "ref_id" keys.
######### OPENING REQUIREMENT #########

The first data_story_sentence must begin with one of: "Across ", "From ", "Behind every ", "Focusing on ", or "In ".
\end{lstlisting}

\subsection{Prompt Used for Alternative Suggestions}
\begin{lstlisting}[basicstyle=\ttfamily\small,
                   breaklines=true,
                   frame=single,
                   backgroundcolor=\color{gray!10},
                   numbers=none]
You are an assistant helping a user reflect on their reasoning as they explore data and build a personal narrative.

Your reflection should help the user:
- Reconsider implicit assumptions
- Explore alternative explanations
- Think about branching to resolve contradictions

You must keep the reflection grounded within the scope of the provided datasets.

############################
RELEVANT DATA SOURCES:
The following dataset categories have been identified as relevant to the current insight. Stay within these domains when suggesting reflection prompts:

${JSON.stringify(firstStepData.related_source, null, 2)}

Only use reasoning related to these domains. Do not speculate outside them.

############################
CONTEXTUAL INSIGHTS:
Here are potentially related or contradictory insights previously written by the user. You may use them to explain reasoning tension or provide contrast.

${JSON.stringify(relatedSentencesContext, null, 2)}

Each item includes:
- node_id
- sentence_id  
- sentence_content
- reason_for_relevance (e.g., "same topic", "opposite claim", "same borough", "time mismatch")

You may reference these insights in your reflection, but only if they meaningfully relate to the current sentence.

############################
INPUT:
{
  "node_id": ${firstStepData.node_id},
  "sentence_id": "${firstStepData.sentence_id}",
  "sentence_content": "${firstStepData.sentence_content}"
}

############################
OUTPUT FORMAT:
Return a single JSON object in this format:
{
  "node_id": ${firstStepData.node_id},
  "sentence_id": "${firstStepData.sentence_id}",
  "sentence_content": "${firstStepData.sentence_content}",
  "reflect": [
    {
      "prompt": "<detailed reflection question>",
      "reason": "<why this prompt is relevant>",
      "related_sentence": {
        "node_id": <number>,
        "sentence_content": "<string>"
      } | null
    }
  ]
}

If you cannot find anything to reflect on, return "reflect": []
\end{lstlisting}

\subsection{Prompt Used for Interaction Capture}
\begin{lstlisting}[basicstyle=\ttfamily\small,
                   breaklines=true,
                   frame=single,
                   backgroundcolor=\color{gray!10},
                   numbers=none]
You are an assistant helping a user write a data-supported personal narrative. The user is exploring a dashboard to decide which borough in London to move to, based on factors like safety, income, housing, education, and quality of life.

They have just interacted with a specific view on the dashboard and clicked "Capture Insight." Your task is to write one concise, readable narrative sentence based on the data they were exploring.

**Your role**: write one concise, data-backed narrative sentence that would be a *natural follow-up* to the user's current sentence, using ONLY the evidence contained in the interaction log.

**Avoid:** Jargon, vague summaries, or speculative statements. Use the actual numbers and trends from the data. The sentence should sound human and fit into a longer narrative paragraph.

############################
INPUT JSON
############################
{
  "narrative_context": <string>,          // a full narration of the user's exploration
  "current_sentence": <string>,           // the sentence they just wrote or are editing
  "interaction_log": [                    // EXACTLY the last 5 dashboard events
    /* ARRAY IS IN CHRONOLOGICAL ORDER:
       index 0 = oldest, index 4 = most recent (most important) */
    {
      "elementId": <string>,              // unique view id
      "elementName": <string>,            // human-readable name
      "elementType": <"chart"|"map" ...>, // what kind of view
      "action": <string>,                 // e.g. "filter_change", "interactive_hover"
      "dashboardConfig": {                // static metadata about the view
        "title": <string>,
        "view_type": <string>,
        "variable_map": { ... }
      },
      "chartData": { ... } or [ ... ]     // numbers or categorical values revealed
    },
    ...
    /* 5 objects total */
  ]
}

**Important reading rules**
1. The **4th and 5th objects (indexes 3 and 4)** represent the user's freshest focus; prefer those for your sentence.
2. Focus on whichever interaction(s) provide a clear, interesting, NUMERIC or CATEGORICAL takeaway.  
   \textbullet  If multiple views are relevant, you may combine two, but keep the sentence short.  
   \textbullet  Ignore hovers that do not reveal new numbers.
3. Do not invent numbers. Quote or paraphrase only what is present.
4. Write in **plain language** (journalistic tone, not code).
5. The sentence should help the user decide where to live (safety, income, diversity, housing, etc.).
6. If NO clear insight is possible, return `"narrative_suggestion": null`.

############################
YOUR SINGLE-LINE JSON OUTPUT
############################
Return a JSON object:

{
  "narrative_suggestion": "<one well-formed English sentence or null>",
  "source_elementId": "<elementId you relied on most>",
  "source_view_title": "<dashboardConfig.title>",
  "explanation": <explanation on why this is a reasonable followup, what insight would it provide in 10-20 words>
}
\end{lstlisting}

\subsection{Prompt Used for Insight Timeline}
\begin{lstlisting}[basicstyle=\ttfamily\small,
                   breaklines=true,
                   frame=single,
                   backgroundcolor=\color{gray!10},
                   numbers=none]
You are an assistant helping a user track how their narrative insights evolve while exploring data.

############################
RELEVANT DATASETS:
The following datasets have been identified as potentially relevant to the current sentence:
${related_datasets ? `
Categories: ${related_datasets.related_categories?.join(', ') || 'None identified'}
Columns: ${related_datasets.related_columns?.join(', ') || 'None identified'}
` : 'No specific datasets identified for this sentence.'}

############################
Your task:
For the given current sentence and the previous one (if available), return a JSON object with:

1. The original \`node_id\` and \`sentence_id\` EXACTLY as provided.
2. A \`changed_from_previous\` object describing what changed, if anything, across time, geography, topic, measure, or logic.
3. A \`related_source\` object containing the relevant categories and columns from the dataset.
4. A \`related_sentence\` object identifying the most relevant previous sentence in the active path, if any exists.

If no previous sentence is provided, set \`changed_from_previous\` to \`null\`.
If no related sentence is found in the active path, set \`related_sentence\` to \`null\`.

############################
Drift types:
1. **Provide Overview** - Establishes the broader context, scope, or big picture within which the story unfolds. Often sets the stage by situating data points relative to larger trends, baselines, or background information.
2. **Adjust** - Changes the framing or focus of the narrative by shifting scale, perspective, or grouping --- e.g., moving from national to local, yearly to monthly, or averages to distributions.
3. **Detect Pattern** - Highlights a relationship, anomaly, or notable trend discovered in the data. These are often the "aha" moments of the story --- where the writer points out evidence that stands out.
4. **Match Mental Model** - Connects the story to prior expectations, domain knowledge, or intuitive explanations. It makes the narrative resonate with what the audience already believes or knows, or challenges those expectations.

Allowed severity: \`none\`, \`minor\`, \`moderate\`, \`critical\`

############################
INPUT DATA:
${JSON.stringify(llmInput, null, 2)}

############################
OUTPUT JSON FORMAT:
[
  {
    "node_id": <number>,
    "sentence_id": "<string>",
    "sentence_content": "<string>",
    "changed_from_previous": {
      "drift_types": [ ... ],
      "severity": "<severity-level>",
      "dimensions": {
        "<what changed>": "<from - to>"
      }
    } | null,
    "related_source": {
      "related_categories": ["category1", "category2"],
      "related_columns": ["column1", "column2"]
    },
    "related_sentence": {
      "node_id": <number>,
      "reason": "<reason why they are related>"
    } | null
  }
]
\end{lstlisting}

\subsection{Prompt Used for Inquiry Labels}
\begin{lstlisting}[basicstyle=\ttfamily\small,
                   breaklines=true,
                   frame=single,
                   backgroundcolor=\color{gray!10},
                   numbers=none]
You are a reasoning assistant enriching a user's inquiry graph for exploratory data analysis. The graph contains only ISSUE nodes (questions). For each issue, add (a) an optional position (likely answer), (b) an optional argument (justification), and (c) ISSUE-ISSUE links. Use ONLY what is present or clearly implied in the user's sentences. Do not speculate.

DEFINITIONS (IBIS-aligned, concise)
- position_suggested_by (optional): The likely or tentative ANSWER to the issue found in the user's text. Include:
  \textbullet  text: short paraphrase or quote of the answer
  \textbullet  confidence: "high" (asserted/verified), "medium" (tentative), "low" (weak hint)
- argument_suggested_by (optional): The JUSTIFICATION supporting that position. Include:
  \textbullet  text: brief summary (data reference, mechanism, pattern, or comparison)
  \textbullet  basis: one of "data" | "mechanism" | "pattern" | "comparison" | "other"
- links (ISSUE-ISSUE only; use sparingly and only if justified):
  \textbullet  suggested_by: Target issue was prompted as a follow-up/side question by this issue.
  \textbullet  generalized_from: Current issue is a broader abstraction of a more specific earlier issue.
  \textbullet  specialized_from: Current issue is a narrower/drilled-down version of a broader earlier issue.
  \textbullet  replaces: Current issue supersedes or rephrases an earlier issue (clearer scope/time/geo).

INPUT
Narrative sentences:
${JSON.stringify(sentenceList, null, 2)}

Extracted issues (from Layer 1):
${JSON.stringify(issueList, null, 2)}
// Each issue has: qid, title, sentenceRefs, status

OUTPUT
Return a JSON array where each item enriches one issue by qid. Do NOT repeat title/status/sentenceRefs. Use null for missing fields.

[
  {
    "qid": "<issue qid>",
    "position_suggested_by": {
      "text": "<answer from user text>",
      "confidence": "low" | "medium" | "high"
    } | null,
    "argument_suggested_by": {
      "text": "<justification from user text>",
      "basis": "data" | "mechanism" | "pattern" | "comparison" | "other"
    } | null,
    "links": [
      {
        "qid": "<target issue qid>",
        "type": "suggested_by" | "generalized_from" | "specialized_from" | "replaces",
        "explanation": "<brief, text-grounded reason>"
      }
    ]
  }
]
\end{lstlisting}

\subsection{Prompt Used for Inquiry Issues}
\begin{lstlisting}[basicstyle=\ttfamily\small,
                   breaklines=true,
                   frame=single,
                   backgroundcolor=\color{gray!10},
                   numbers=none]
You are an assistant that analyzes the user's narrative path during exploratory analysis and identifies which sentences raise meaningful questions (issues) worth tracking.

These issues may be:
- Explicit questions (e.g. "Did X happen in 2022?")
- Implicit inquiries (e.g. "I should check whether..." or "It seems like...")
- Investigative goals that signal something unresolved or still unfolding

Each issue should:
- Be framed as a **question** (`title`)
- Be assigned a **status**:
   - `"open"` if the question is raised but unanswered
   - `"resolved"` if the narrative includes clear follow-up or confirmation
   - `"stalled"` if the question was raised earlier but left unresolved and later abandoned or deferred (not in the final sentences)
- Be grounded in one or more **sentenceRefs** (e.g., `["s3"]`)

############################
INPUT USER NARRATIVE:
${JSON.stringify(sentenceList, null, 2)}

Each sentence item contains:
- sentence_id
- content

############################
OUTPUT FORMAT:
Return a JSON array of identified issues:
[
  {
    "qid": "<unique string ID, prefixed with iss_>",
    "title": "<question title>",
    "status": "open" | "resolved" | "stalled",
    "sentenceRefs": ["<sentence_id>", ...]
  }
]

If no issues are identified, return an empty array: `[]`

\end{lstlisting}







\end{document}